\documentclass[aps,showpacs,amsmath,amssymb]{revtex4}
\usepackage{graphicx}

\usepackage{bm}
\usepackage{float}
\usepackage{mathtools}
\usepackage{multirow}
\usepackage{commath}
\usepackage{braket}
\usepackage{subfig}
\usepackage{bbold}
\usepackage{amsthm}
\usepackage{mathrsfs}
\usepackage[usenames]{xcolor}
\usepackage{gensymb}
\usepackage{appendix}
\usepackage{graphicx}

\usepackage[normalem]{ulem}

\DeclareMathOperator{\Tr}{Tr}

\begin{document}
\def\mean#1{\left< #1 \right>}

\title{Stochastic thermodynamics of quantum maps with and without equilibrium}

\author{Felipe Barra\footnote{fbarra@dfi.uchile.cl} and Crist\'obal Lled\'o}
\affiliation{Departamento de F\'isica, Facultad de Ciencias F\'isicas y Matem\'aticas, Universidad de Chile, Santiago, Chile}

\begin{abstract}
We study stochastic thermodynamics for a quantum system of interest whose dynamics are described by a completely positive trace-preserving (CPTP) map as a result of its interaction with a thermal bath. 
We define CPTP maps with equilibrium as CPTP maps with an invariant state such
that the entropy production due to the action of the map on the invariant state vanishes. Thermal maps are a subgroup of CPTP maps with equilibrium.
In general, for CPTP maps,
the thermodynamic quantities, such as the entropy production or work performed on the system, depend on the combined state of the system plus its environment. We show that these quantities can be written in terms
of system properties for maps with equilibrium. The relations that we obtain are valid for arbitrary coupling strengths between the system and the thermal bath.   
The fluctuations of thermodynamic quantities are considered in the framework of a two-point measurement scheme. We derive the entropy production fluctuation theorem for general maps and a fluctuation relation for the stochastic work on a system that starts in the Gibbs state. Some simplifications for the probability distributions in the case of maps with equilibrium are presented.
We illustrate our results by considering spin 1/2 systems under thermal maps, non-thermal maps with equilibrium, maps with non-equilibrium steady states and concatenations of them. 
Finally, we consider a particular limit in which the concatenation of maps generates a continuous time evolution in Lindblad form for the system of interest, and 
we show that the concept of maps with and without equilibrium translates into Lindblad equations with and without quantum detailed balance, respectively. The consequences for the thermodynamic quantities in this limit are discussed.
\end{abstract}
 
\pacs{
05.70.Ln,  
05.70.-a,   
03.65.Yz 
75.10.Pq 
}
\maketitle

\section{Introduction}

At present, experiments are at the edge of dealing with thermal machines
where quantum mechanics should be relevant; therefore, issues such as the manifestations of non-classical features in their behavior and a proper thermodynamic formulation for quantum machines are currently being investigated~\cite{KoslovPRX}. For various systems of interest, the interaction energy between the system and the environment can be neglected compared
to the energy of the system and the energy of the bath. In this case, a thermodynamic framework referred to as ``in the weak coupling" is very successful~\cite{koslov-entropy,BrandnerSeifert}. In particular, work is performed on the system by externally varying a control parameter of the system Hamiltonian, for instance, by changing a field that raises an energy level of the system.  

For other systems of interest, one can engineer the coupling between the system and external probes, making it global or local and switching it on and off in a controlled manner; see, e.g.,~\cite{ion-trap}.
In this paper, we consider such a setup, and with the purpose of studying thermodynamic processes, we consider the simple situation in which the external probe is prepared in a thermal state. Thus, we shall refer to the probe as the bath, even though we are not assuming that the probe is macroscopic.
The experimenter controls the coupling at some work cost, and a certain amount of heat will also flow between the system and the bath. These thermodynamic quantities are completely determined by the coupling energy in this case.  
An appropriate formalism for analyzing these quantum evolutions and the thermodynamic 
behavior is that of completely positive trace-preserving (CPTP) maps~\cite{MHPPRE2015, Goold}.
In particular, one would like to account for the coupling between the system and the bath non-perturbatively. The thermodynamic properties of systems strongly coupled to a bath are not well understood, although they have been
considered in~\cite{seifert-strong,jarzynski-strong,esposito-NJP,anders-strong}.

We study quantum stochastic thermodynamics as formulated in~\cite{esposito-NJP,Reeb-Wolf,HPNJP2013}, where one considers that the system plus environment evolve unitarily during the process from an uncorrelated initial state; therefore, the change in the system due to this process is given by a CPTP map. In this formulation, the strength of the coupling is arbitrary, but the thermodynamic quantities, such as the work performed on the system, the total heat exchange between the system and the bath and the total entropy production, are non-local quantities that are expressed in terms of the total system-bath density matrix at the beginning and end of the process. In the limit where the strength of the coupling vanishes, one recovers
the expressions obtained in the weak coupling~\cite{BreuerBook,HSJMP1978,SLACP1978,Alicki}, which are local, i.e., they depend only on system operators. In our study, we go beyond the analysis of averaged quantities and also consider their stochastic versions defined using a two-point measurement scheme~\cite{esposito-mukamel-RMP}. This stochastic thermodynamics relies on the concept of stochastic trajectory. The thermodynamic quantities are defined for every trajectory in such a way that upon
averaging, the standard definitions for the mean quantities are recovered. Fluctuation theorems reveal the statistical properties of these quantities for arbitrary non-equilibrium processes~\cite{Gallavotti-cohen, KurchanJPA, LebSphon, Crooks1, Gasp, SeifertPRL05,experimentales}, and they have been established for classical Hamiltonians or stochastic systems~\cite{Gallavotti-cohen,LebSphon}, isolated quantum systems~\cite{Kurchan, Tasaki, Gaspard, Campisi} and for CPTP maps~\cite{JHPRE2012, HPNJP2013,MHPPRE2015} representing a system driven by a time-dependent
protocol. Here, we derive these equalities but emphasize a process driven by a controlled interaction with the bath. 

A process may involve several interactions with the bath~\cite{anders-giovannetti}; thus, a sequence of different maps may act on the system. In particular, repeating the same sequence, i.e., a periodic driving, may bring the system to a stationary or invariant state. 
In general, an invariant state of a CPTP map represents a non-equilibrium steady state (NESS) in the sense that the entropy production due to the action of the map over the state is strictly positive. We call such states {\it maps with NESS}. We define {\it maps with equilibrium} as maps in which the entropy production associated with the action of the map on the invariant state vanishes. \textit{Thermal maps} are a subgroup of maps with equilibrium, and they have  the canonical thermal state of the system as an invariant state. We will provide further details later.
From the perspective of thermodynamics, maps with equilibrium have an interesting property in that the thermodynamic quantities depend only on the system variables, even in the strong-coupling regime. These maps generalize thermal maps, and they are related to the existence of conserved quantities. 
We obtain analytical expressions for the thermodynamic quantities and, in particular situations, for the stochastic quantities. We illustrate the relevance of our results in spin 1/2 systems under the evolution of a thermal map, a non-thermal map with equilibrium and maps with NESS. 

Lindblad master equations are an important tool for studying open quantum systems. Systems that are weakly and passively coupled to a heat bath are described by a Lindblad equation with Lindblad operators, which are eigenoperators of the system Hamiltonian~\cite{BreuerBook}. 
In this case, the evolution satisfies the condition of quantum detailed balance~\cite{HSJMP1978,SLACP1978} with respect to the Gibbs thermal state, which allows a consistent thermodynamics formulation~\cite{Alicki}. 
Since any Lindblad equation 
preserves the positivity and the trace of the density matrix, it is natural to attempt to extend the previous scenario to other choices for the Lindblad operators. For instance, in the so-called
boundary-driven Lindblad equations~\cite{prosenXY,Wichterich07}, the dissipator acts on the boundaries of the open system with Lindblad operators that are not eigenoperators of the system Hamiltonian.
If one is interested in thermodynamic processes in these systems, as in ~\cite{linden,AB2013,Shigeru,Bojan,norditaUs}, one has to be careful with the definitions 
of quantities such as work, heat and entropy production~\cite{kosloff} because one cannot infer them from sole knowledge of the Lindblad equation.
Since Lindblad master equations can be obtained from the repeated concatenation of CPTP maps in a particular limit~\cite{Attal,Karevski}, our results provide a proper thermodynamic description
of the processes described by the corresponding Lindblad dynamics. We show that maps with and without equilibrium translate into Lindblad equations with and without quantum detailed balance, respectively. 
Thermal maps generate a Lindblad dynamics with quantum detailed balance with respect to the Gibbs thermal state, whereas a generic map with equilibrium generates a Lindblad dynamics with quantum detailed balance with respect to an equilibrium state that may not be the Gibbs state. Maps with NESS generate Lindblad equations without quantum detailed balance. 
The consequences for the thermodynamic quantities are discussed, and we particularly emphasize the work that accompanies a process whose dynamics is generated by a time-independent Lindblad master equation with a non-Gibbsian equilibrium state or with a NESS.

The remainder of this article is organized as follows. In section \ref{traj.map.sec}, we introduce stochastic thermodynamics for CPTP maps, we define CPTP maps with equilibrium, and we study their main properties. Then, we prove a fluctuation theorem for the entropy production
and a work fluctuation theorem similar to the Crooks fluctuation theorem~\cite{Crooks1} for CPTP maps with and without equilibrium. In section \ref{sec.aps}, we apply our results to spin 1/2 systems. Subsequently, in section \ref{sec: Lindblad}, we consider the limit in which the concatenation of maps provides a Lindblad dynamics and discuss the consequences for the thermodynamic properties of the systems.
We conclude this article in section \ref{secCONC}.

\section{Stochastic thermodynamics for completely positive trace-preserving maps}
\label{traj.map.sec}

In this section, we present the main results of stochastic thermodynamics for an
open quantum system whose dynamics is controlled by a CPTP map. Consider a system and a bath that have a joint evolution governed by the unitary  $U=\mathcal{T}e^{-i\int_0^\tau(H_S(t)+H_B+V(t))dt}$ ($\hbar =1$ throughout the text), where $\mathcal T$ is the time ordering operation.
The Hamiltonian $H_B$ of the heat bath is constant in time. The coupling between the system and the bath is represented by an interaction energy $V(t)$ that vanishes for $t<0$ and $t>\tau$. During the joint evolution, the system can be driven in a cycle, i.e., its Hamiltonian may be time dependent $H_S(t)$, but $H_S(0)=H_S(\tau)=H_S$. We consider this condition because we are interested
in systems that have an invariant state, and an arbitrarily driven system will generally not have such a state. Later, we will consider examples with constant Hamiltonians $H_S$.
We assume that $H_B$ and $H_S$ are non-degenerate. 
The eigenstates of the system are in the Hilbert space ${\mathcal H}_S$, and those of the heat bath are in ${\mathcal H}_B$. Initially, at time $t=0$, 
the system and bath are uncoupled, i.e., their density matrix is the tensor product of the respective density matrices  $\rho_{\rm tot}=\rho_S\otimes \omega_\beta(H_B)$, where $\omega_\beta(H_B)=\frac{e^{-\beta H_B}}{Z_B}$ is the canonical thermal state of the bath with $\beta=T^{-1}$ ($k_B=1$), the inverse temperature of the bath, and $Z_B={\rm Tr} \,e^{-\beta H_B}.$
After a lapse of time $\tau$ in which the system and bath are coupled, the initial state $\rho_{\rm tot}$ in the product Hilbert space ${\mathcal H}_S\otimes {\mathcal H}_B$ changes to a new state,
\begin{equation}
\label{unitary}
\rho'_{\rm tot}=U \left(\rho_S\otimes\omega_\beta(H_B)\right) U^\dag.
\end{equation}
In the following, we denote $\rho_S'={\rm Tr}_B\rho_{\rm tot}'$ and $\rho_B'={\rm Tr}_S\rho_{\rm tot}'$, where ${\rm Tr}_X$ is the partial trace over subsystem $X$. By tracing out the bath, one obtains a completely positive trace-preserving (CPTP) map $\mathcal{E}$ for the system evolution
\begin{equation}
\rho'_S=\mathcal{E}(\rho_S)={\rm Tr}_B\left[U \left(\rho_S\otimes\omega_\beta(H_B)\right) U^\dag\right]=\sum_{ij} M_{ij}\rho_S M_{ij}^\dag
\label{CPTP}
\end{equation}
with Kraus operators
\begin{equation}
M_{ij}=\sqrt{\frac{e^{-\beta \varepsilon_i}}{Z_B}}\langle j|U|i\rangle.
\label{kraussHeat}
\end{equation} 
Here, $|i\rangle$ and $|j\rangle$ are the eigenstates of the bath Hamiltonian $H_B$ with eigenvalues $\varepsilon_i$ and $\varepsilon_j$, respectively. Note that as required for the trace preservation of $\mathcal{E}$
\[
\sum_{ij}M^\dag_{ij}M_{ij}=\sum_{i}\frac{e^{-\beta \varepsilon_i}}{Z_B}\langle i|U^\dag\sum_j|j\rangle\langle j|U|i\rangle=\sum_{i}\frac{e^{-\beta \varepsilon_i}}{Z_B}\langle i|I_S\otimes I_B|i\rangle=I_S.
\]
As is well known, there are many choices for the set of Kraus operators $M_{ij}$ that produce the same map $\mathcal{E}$.
This particular representation provides a relation between the evolution of the system and the changes in the bath. The quantum operation $M_{ij}\rho_S M_{ij}^\dag$ provides the change in the system associated with the transition $|i\rangle\to|j\rangle$ in the bath.
For a single map, we define and study the relevant quantities for stochastic thermodynamics. This will be generalized later for sequences of maps.

\subsection{Quantum trajectories}

In the context of stochastic thermodynamics, one introduces the fluctuating quantities as the result of a two-point measurement~\cite{esposito-mukamel-RMP}. These measurements induce a probability distribution for the possible outputs of the thermodynamic quantities characterizing the process associated to $\mathcal E$.
Initially, with the system in an arbitrary state $\rho_S$, a non-selective projective measurement of a non-degenerate system operator $A$ is performed,  leaving the system in the state
$\bar{\rho}_S=\sum p_i(n) |a_{n}\rangle\langle a_{n}|$ with $p_i(n)={\rm Tr}_S(\ket{a_n}\bra{a_n}\rho_S)$. The energy $H_B$ is measured on the bath that was initially in a thermal state, leaving it in the state $\ket{i}\bra{i}$ with probability $e^{-\beta \epsilon_i}/Z_B$. The uncorrelated state $\bar{\rho}_S\otimes\ket{i}\bra{i}$ evolves unitarily, and at the end of the process, the bath Hamiltonian $H_B$ is measured again. The bath is then left in state $\ket{j}\bra{j}$. The pair $k=(ij)$ identifies a Kraus operator $M_{ij}$, and one can observe that the probability of measuring the bath energies $\varepsilon_i$ initially and $\varepsilon_j$ finally is given by $p_k(\bar\rho_S)={\rm Tr}[\mathcal{E}_k(\bar\rho_S)]$, where $\mathcal{E}_k(\cdot)\equiv M_k\cdot M_k^\dag$. Thus, if one is concerned about the system, one can say that during its evolution, one of the operations $\mathcal{E}_k(\cdot)$ occurs with probability 
$p_k(\bar\rho_S)$,
but if no register is kept of the value $k$, then the final state is $\rho'_S=\sum p_k(\bar\rho_S) \rho'_k=\mathcal{E}(\bar\rho_S)$ with $\rho'_k=\mathcal{E}_k(\bar\rho_S)/p_k(\bar\rho_S)$.
Additionally, at the end of the process, a non-selective projective measurement of another non-degenerate system operator $B$ is performed, and finally, the system is in state $\bar\rho_S'=\sum p_f(n)|b_{n}\rangle\langle b_{n}|$ with $p_f(n)={\rm Tr}[\mathcal{E}(\bar\rho_S) |b_{n}\rangle\langle b_{n}|]$.
A {\it trajectory}  is defined as the sequence of values $a_n,k,b_m$ and is denoted by $\gamma=\{n,k,m\}$. Its probability $p(\gamma)=p(m,k|n)p_i(n)$ 
is the probability of measuring $b_m$ after the operation $\mathcal{E}_k$ occurs, given that the system was initially in the state $|a_{n}\rangle\langle a_{n}|$, times the probability of the latter, i.e.,
\begin{equation}
p(\gamma)=|\langle b_m|M_k|a_n\rangle|^2 p_i(n).
\label{prob1}
\end{equation}
Note that $\sum_{k,n}p(m,k|n)p_i(n)=p_f(m)$. The probability $p(\gamma)$ that we obtained with these non-selective measurements corresponds to the probability one would compute in practice by repeating many times the (identically prepared) experiment with selective measurements. 
Explicitly, with the expression in Eq.~(\ref{kraussHeat}) for the Kraus operators, the probability of a trajectory $\gamma=\{n,k,m\}=\{n,(ij),m\}=\{n,i;m,j\}$ is, according to Eq.~(\ref{prob1}),
\begin{equation}
p(\gamma)=|\langle j,b_m|U|i,a_n\rangle|^2 \frac{e^{-\beta \varepsilon_i}}{Z_B}p_i(n).
\label{psm1}
\end{equation}

\subsection{Stochastic thermodynamics}
\label{stoch.term.sec}

In the previous subsection, we obtained the probability $p(\gamma)$ of the quantum trajectory $\gamma$ for a quantum system interacting with a bath.
We now associate the stochastic thermodynamic quantities of these trajectories.

If one measures the energy at the beginning and at the end of the process, i.e., $A=B=H_S$, we have the stochastic system energy change $\Delta e_\gamma=\epsilon_m-\epsilon_n$. Here, $\epsilon_n$ denotes the eigenvalue of $H_S$ associated with the eigenvector $\ket{\epsilon_n}$,
and in this case, $\ket{a_n}=\ket{b_n}=\ket{\epsilon_n}$. 
The stochastic heat flow to the system $q_\gamma$ associated with the trajectory $\gamma$ corresponds to the negative energy change of the bath, and we assume that it is
obtained knowing $k$. Indeed, the trajectory $\gamma=\{n,k,m\}=\{n,i;m,j\}$, with probability given in Eq.~(\ref{psm1}), represents
the transition $|i\rangle\to|j\rangle$ in the bath whose energy change $\varepsilon_j-\varepsilon_i$ is minus the stochastic heat flow to the system, i.e., $q_\gamma=\varepsilon_i-\varepsilon_j$.
According to the first law of stochastic thermodynamics, the stochastic work is given by
\begin{equation}
w_\gamma=\Delta e_\gamma-q_\gamma.
\label{w-gamma}
\end{equation}
These fluctuating quantities will be studied through their distribution, for instance, for heat and work
\begin{equation}
p(q)=\sum_\gamma \delta(q-q_\gamma)p(\gamma),\quad p(w)=\sum_\gamma \delta(w-w_\gamma)p(\gamma).
\label{p(w)}
\end{equation}

By defining the averages over the trajectories  $\Delta E=\sum_\gamma \Delta e_\gamma p(\gamma)$, $Q=\sum_\gamma q_\gamma p(\gamma)$ and $W=\sum_\gamma w_\gamma p(\gamma)$, one obtains
\begin{equation}
\Delta E={\rm Tr}[H_S(\bar\rho'_S-\bar\rho_S)]={\rm Tr}[H_S(\rho'_S-\rho_S)]
\label{Av.Energy}
\end{equation}
\begin{equation}
Q={\rm Tr}[H_B(\omega_\beta(H_B)-\rho'_B)]
\label{Av.Heat}
\end{equation}
and
\begin{equation}
W={\rm Tr}[(H_S+H_B)(\rho'_{\rm tot}-\rho_{\rm tot})],
\label{Av.Work}
\end{equation} 
satisfying the first law $\Delta E=W+Q$.

Although measuring density matrices is highly non-trivial from an experimental perspective, if one measures $A=\rho_S$ and $B=\rho'_S$, given in Eq.~(\ref{CPTP}), one obtains the stochastic entropy change $\Delta s_\gamma=-\ln p_f(m)+\ln p_i(n)$. Note that in this case, $\bar\rho_S=\rho_S$ and $\bar\rho_S'=\rho_S'$. As the environment consists of a heat bath with inverse temperature $\beta$, the stochastic entropy flow $\beta q_\gamma$  
and the entropy change $\Delta s_\gamma$ define the stochastic entropy production
\begin{equation}
\Delta_i s_\gamma=\Delta s_\gamma-\beta q_\gamma. 
\label{epgamma}
\end{equation}
This fluctuating quantity will be studied through the entropy production distribution
\begin{equation}
p(\Delta_is)=\sum_\gamma \delta(\Delta_is-\Delta_is_\gamma)p(\gamma).
\label{p(si)}
\end{equation}

These definitions are such that upon averaging over the trajectories, one obtains 
$\Delta_i S\equiv\sum_\gamma \Delta_i s_\gamma p(\gamma)=\Delta S-\beta Q$,
with 
\begin{equation}
\Delta S=-{\rm Tr}[\rho_S'\ln\rho_S']+{\rm Tr}[\rho_S\ln\rho_S]
\label{Av.Ent}
\end{equation}
the von Neumann entropy change and $Q$ given in Eq.~(\ref{Av.Heat}). The averaged entropy production $\Delta_iS$ can be expressed as
\begin{equation}
\Delta_iS=D(\rho_{\rm tot}'||\rho_S'\otimes \omega_\beta(H_B))\geq 0,
\label{Av.Ent.Prod}
\end{equation}
where $D(a||b)=\Tr[a\ln a] - \Tr[a \ln b].$ 
These averaged expressions are valid beyond the two-point measurement scheme that we consider here and were first obtained in~\cite{esposito-NJP}.
Therefore, we consider that for a process $\rho_S\to\rho_S'={\mathcal E}(\rho_S)$, the averages simultaneously satisfy $\Delta E=W+Q$ and $\Delta_i S=\Delta S-\beta Q$
with the quantities given in Eqs.~(\ref{Av.Energy},~\ref{Av.Heat},~\ref{Av.Work},~\ref{Av.Ent}, and~\ref{Av.Ent.Prod}), even though the fluctuations of the entropy production and the work can be studied simultaneously only if $[H_S,\rho_S]=[H_S,\rho_S']=0$. Note that for their evaluation, particularly for the work, Eq.~(\ref{Av.Work}), and entropy production, Eq.~(\ref{Av.Ent.Prod}), we need to know the full state $\rho_{\rm tot}'$. In contrast, in the weak-coupling limit, where $V(t)$ can be neglected in comparison to $H_S$ and $H_B$, the thermodynamic quantities depend only on the states $\rho_S'$ and $\rho_S$ of the system of interest. We will subsequently show that this simplification can occur for the strongly coupled systems defined below.

\subsection{ Maps with thermodynamic equilibrium}
\label{sec.map.eq.}
Let us assume that the map ${\mathcal E}$ has an attractive invariant state $\pi$ defined as $\lim_{N\to\infty}{\mathcal E}^N(\rho_S)=\pi$ $\forall \rho_S$ and $\pi={\mathcal E}(\pi)$.
An invariant state is thermodynamically characterized by $\Delta S=0=\Delta E$, as shown in Eq.~(\ref{Av.Energy}) and Eq.~(\ref{Av.Ent}). We will say that this invariant state is an {\it equilibrium state} if $\Delta_iS=0$, i.e., if the entropy production, Eq.~(\ref{Av.Ent.Prod}), vanishes by the action of $\mathcal E$ on $\pi$. Maps with these special states are called maps with equilibrium. If the entropy produced by the action of the map ${\mathcal E}$ on $\pi$ provides $\Delta_iS>0$, then we say that the invariant state is a non-equilibrium steady state. In this case, one obtains $Q=-\beta^{-1}\Delta_iS<0$ and $W=\beta^{-1}\Delta_iS>0$. This means that the non-equilibrium steady state is sustained by the work performed by an external agent implementing the map on the system, which is dissipated as heat. In this situation, we say that ${\mathcal E}$ is a map with NESS.

According to Eq.~(\ref{Av.Ent.Prod}), $\Delta_iS=0$ for the steady state $\pi$ if and only if $\pi\otimes\omega_\beta(H_B)=U \left(\pi\otimes\omega_\beta(H_B)\right) U^\dag$.  
Equivalently, 
if the unitary $U$  in Eq.~(\ref{unitary}) satisfies $[U,H_0+H_B]=0$, where $H_0$ is an operator in the Hilbert space of the system ${\mathcal H}_S$, then the product state $\omega_\beta(H_0)\otimes\omega_\beta(H_B)$, with $\omega_\beta(H_0)=\frac{e^{-\beta H_0}}{Z_0}$, where $Z_0={\rm Tr}[e^{-\beta H_0}]$, is invariant under the unitary evolution in Eq.~(\ref{unitary}) and $\omega_\beta(H_0)$ is an equilibrium state for the map in Eq.~(\ref{CPTP}).

It follows from $[U,H_0+H_B]=0$ that \begin{equation}
Q={\rm Tr}[H_B(\omega_\beta(H_B)-\rho_B')]={\rm Tr}[H_0(\rho_S'-\rho_S)].
\label{Eq.prop}
\end{equation}
Thus, for maps with equilibrium, the average work, Eq.~(\ref{Av.Work}), simplifies to
\begin{equation}
W={\rm Tr}_S[(H_S-H_0)(\rho_S'-\rho_S)]
\label{Av.Work.Eq}
\end{equation}
which is determined by the system state only. 

When the map has an equilibrium state $\omega_\beta(H_0)$, the entropy production also reduces to an expression that does not involve the state of the bath.
Indeed, from $\Delta_i S=\Delta S-\beta Q$ with Eq.~(\ref{Eq.prop}), we obtain $\Delta_iS={\rm Tr}[\rho_{S}\ln \rho_{S}]-{\rm Tr}[\rho_{S}'\ln \rho_{S}']-{\rm Tr}[(\rho_S-\rho_S')\ln \omega_\beta(H_0)]$, which can be rearranged into
\begin{equation}
\Delta_iS
=D(\rho_S||\omega_\beta(H_0))-D(\rho_S'||\omega_\beta(H_0)),
\label{epthermal}
\end{equation}
which is positive due to the contracting character of the map~\cite{BreuerBook}.
Note that Eqs.~(\ref{Eq.prop}), (\ref{Av.Work.Eq}) and (\ref{epthermal}) are exact for maps with an equilibrium state and do not require a weak-coupling ($V(t)$ small) condition to be satisfied.

Let us consider in further detail two particular cases of interest: 

{\it i) Stochastic thermodynamics of thermal maps:}
If $H_0=H_S$, then the map is called thermal~\cite{terry1,terry2,terry3,Oppenheim}.
The equilibrium state of thermal maps is the Gibbs thermal state $\omega_\beta(H_S)=e^{-\beta H_S}/Z_S$ with $Z_S={\rm Tr}[e^{-\beta H_S}]$. 
For thermal maps, the average work vanishes for every initial state $\rho_S$, as follows from Eq.~(\ref{Av.Work.Eq}). 
If a thermal map brings the system to the equilibrium Gibbs state, then the entropy production Eq.~(\ref{epthermal}) reduces to the well-known expression~\cite{DeffnerPRL}
$\Delta_iS=D(\rho_S||\omega_\beta(H_S)).$ This is the dissipation that occurs in the relaxation process $\rho_S\to\omega_\beta(H_S).$

Let us now discuss the work fluctuations.  
Consider an energy measurement $H_S$ at the beginning and at the end of a process realized by a thermal map. 
From $[U,H_S+H_B]=0$, it follows that 
if the transition is possible ($p(\gamma)\sim|\langle \epsilon_m j|U|\epsilon_n i\rangle|^2\neq 0$), then it conserves the energy $\epsilon_m+\varepsilon_j=\epsilon_n+\varepsilon_i$.  Therefore, for a thermal map, Eq.~(\ref{w-gamma}) provides   
$w_\gamma=0$ 
for any trajectory $\gamma$ with $p(\gamma)\neq 0$ and
for any initial state $\rho_S$. 
Since thermal maps do not require an external agent who performs or extracts work from the system, they are supposed to describe the passive coupling between a system and a heat bath.

\vspace{0.2cm}

{\it ii) Stochastic thermodynamics of  non-thermal maps with equilibrium:}

We consider maps with an equilibrium state $\omega_\beta(H_0)$ with $H_0\neq H_S$
but restrict ourselves to the particular situation in which $[H_0,H_S]=0$. This is the situation that we will encounter in the examples, and we argue that it will often be the case if the system Hamiltonian is constant during the evolution, i.e., $H_S(t)=H_S\,\forall  t$, and
the coupling is a step function, i.e., $V(t)=V_{\rm loc}$ if $0<t<\tau$ and $V(t)=0$ elsewhere. Indeed, in this case, one finds that the equilibrium condition $[U,H_0+H_B]=0$ implies that $[H_S,H_0]=0$, with the exception of some very particular cases in which $[V,H_0+H_B] = [H_0,H_S]\otimes\mathbb{1}_B$.
Regardless, if $[H_0,H_S]=0$, the first interesting observation is that one can manipulate Eq.~(\ref{epthermal}) and rewrite it as
\[
\Delta_iS=D(\rho_S||\omega_\beta(H_S))-D(\rho_S'||\omega_\beta(H_S))+\beta W,
\]
where $W$ is given in Eq.~(\ref{Av.Work.Eq}). 
This expression coincides with the one derived in~\cite{DeffnerPRL} on thermodynamical grounds for a system weakly coupled to a bath under a cyclic driving ($H_S(0)=H_S(\tau)=H_S$) as those that we consider. Interestingly, if $[H_0,H_S]=0$, then the work fluctuations also become a system property, i.e., one does not need to perform a 
measurement in the bath. Indeed, let us consider the work distribution Eq.~(\ref{p(w)}). Since $[H_0,H_S]=0$ and $H_S$ is non-degenerate, $\ket{\epsilon_n}$ is also an eigenvector of $H_0$, i.e., $H_0\ket{\epsilon_n}=\epsilon_n^0\ket{\epsilon_n}$. Then, from the equilibrium property $[U,H_0+H_B]=0$, one finds that $\epsilon^0_m+\varepsilon_j=\epsilon^0_n+\varepsilon_i$ if $|\langle \epsilon_m,j|U|\epsilon_n,i\rangle|^2\neq 0$, which is proportional to $p(\gamma)$ in Eq.~(\ref{p(w)}).
Therefore, one is allowed to replace the stochastic work Eq.~(\ref{w-gamma}) inside the delta function by $w_\gamma=\epsilon_m-\epsilon_m^0-(\epsilon_n-\epsilon_n^0)$, obtaining
\begin{equation} \label{ecc:dist trabajo eq. map}
p(w)=\sum_{n,m} \delta(w-[(\epsilon_m-\epsilon_m^0)-(\epsilon_n-\epsilon_n^0)])\bra{\epsilon_m}{\mathcal E}(\ket{\epsilon_n}\bra{\epsilon_n})\ket{\epsilon_m}p_i(n),
\end{equation}
completely determined by system quantities and the map ${\mathcal E}$ without the need for measuring the bath. By averaging $w$ with Eq.~(\ref{ecc:dist trabajo eq. map}), we recover an expression that is apparently different from Eq.~(\ref{Av.Work.Eq}), namely, $\Tr[(H_S-H_0)(\rho_S'-\bar \rho_S)]$. However, since the non-selectively measured $\bar \rho_S$ is diagonal in the eigenbasis of $H_S$ (and $H_0$), then it is the same as Eq.~(\ref{Av.Work.Eq}).

Moreover, because $H_S$ and $H_0$ share an eigenbasis, we can simultaneously study the work and entropy production fluctuations in the equilibrium state. One finds here that $p(\Delta_is)=\delta(\Delta_is)$. In fact, using a similar argument as the one used for the work distribution, one can replace $\Delta_is_\gamma$ by $\ln p_i(n)-\ln p_f(m)-\beta (\epsilon^0_m-\epsilon^0_n)$ inside the delta function in Eq.~(\ref{p(si)}). However, since $p_i(n)=e^{-\beta\epsilon_n^0}/Z_0$ and $p_f(m)=e^{-\beta\epsilon_m^0}/Z_0$, one obtains $p(\Delta_is)=\delta(\Delta_is)$, a result also valid for thermal maps ($H_0=H_S$).
In contrast, the equilibrium work fluctuations, Eq.~(\ref{ecc:dist trabajo eq. map}) with $p_i(n)=e^{-\beta\epsilon_n^0}/Z_0$, gives $p(w)\neq \delta(w)$ (although $W=0$) for non-thermal maps with equilibrium while $p(w)= \delta(w)$ for thermal maps.

Let us summarize the results of this subsection. For a map with an equilibrium state, the average thermodynamic quantities can be written in terms of the properties of the system of interest only. 
In general, an active external agent has to provide (or extract) work to perform the map on a state $\rho_S$. Only for thermal maps is $W=0$, and the agent is passive. If the system is in the equilibrium state
$\omega_\beta(H_0)$, the map can be performed with $W=0$, see Eq.~(\ref{Av.Work.Eq}), and $\Delta_iS=0$. In this equilibrium state, the entropy production does not fluctuate $p(\Delta_i s)= \delta(\Delta_is)$, but the work may still present fluctuations $p(w)\neq \delta(w)$ except for thermal maps.
In the general case of a map with equilibrium in which $[H_0,H_S]\neq0$,  we cannot discuss work and entropy production fluctuations simultaneously. If we consider the latter, we still have that the entropy production is a non-fluctuating quantity in the equilibrium state. This can also be obtained as a consequence of the integral fluctuation theorem, as we will see below.

\subsection{Fluctuation Theorems for the Entropy Production}

A central result from stochastic thermodynamics is the detailed fluctuation theorem for the stochastic entropy production $\Delta_is$.
Consider the probability distribution, Eq.~(\ref{p(si)}), of a given $\Delta_is$ value obtained according to the 
$A=\rho_{S}$ and $B=\rho_S'$ two-point measurement procedure. The detailed fluctuation theorem for the entropy production is
\begin{equation}
\ln\frac{p(\Delta_is)}{\tilde p(-\Delta_is)}=\Delta_is,
\label{ep.fluct}
\end{equation}
where $\tilde p(\Delta_is)$ refers to the distribution of the entropy production in a reverse process to be specified later. This equality is derived from the time reversal properties of the system. For the type of systems that we study, it was derived in~\cite{HPNJP2013} for driven Hamiltonians. 
Here, we derive it to emphasize that the fluctuation theorem is also valid for systems driven by other mechanisms, for instance, by the coupling to the bath.

The time reversal operator~\cite{Haake} $\Theta$ is anti-unitary, i.e., $\Theta i=-i\Theta $, and $\Theta ^\dag=\Theta ^{-1}$. 
This operator is defined in the full Hilbert space ${\mathcal H}_S\otimes{\mathcal H}_B$ and is of the form $\Theta_S\otimes\Theta_B$.
The unitary evolution $U$ depends on the time dependence of the system Hamiltonian, but as we mentioned, it can be constant in time, and on the time dependence of the coupling that at least is switched on and off. This time dependence is referred to as the protocol.
If one performs the protocol in the time-inverted sequence, i.e., one considers $\{H_S(\tau-t),V(\tau-t)\}$, the unitary dynamics will be called $\tilde U$. The micro-reversibility principle for non-autonomous systems~\cite{Gasp,Campisi}
relates the forward and backward dynamics by $\Theta ^\dag \tilde U \Theta =U^\dag$. 
Thus, if the unitary operator maps $|\phi\rangle$ to $|\phi'\rangle$, i.e., $|\phi'\rangle=U|\phi\rangle$, the time-reversed state $\Theta |\phi'\rangle$ is mapped to the time-reversed state $\Theta |\phi\rangle$ by the 
time-reversed unitary $\tilde U$. We denote reverse states as $|\tilde\cdot\rangle=\Theta |\cdot\rangle$. We remark that the anti-linearity of $\Theta$ implies that $\bra{\tilde \cdot} = \bra{\Theta\, \cdot} \neq \bra{\cdot}\Theta^\dag$.

 For the time-reversed dynamics, we consider an arbitrarily chosen initial state for the system, $\tilde\rho_S=\sum \tilde p_f(m)|\tilde b_{m}\rangle\langle \tilde b_{m}|$, and a time-reversed thermal state for the bath, $\tilde\rho_B=\sum \frac{e^{-\beta \varepsilon_j}}{Z_B}|\tilde j\rangle\langle \tilde j|$.
Then, the system's time-reversed map $\tilde{\mathcal{E}} (\tilde{\rho}_S) = \tilde{\rho}_S'$ has the representation ${\tilde{\mathcal E}}(\cdot)=\sum_{ij}\tilde M_{ji}\cdot \tilde M_{ji}^{\dag}$ in terms of reversed Kraus operators
\begin{equation}
\tilde M_{ji}=\sqrt{\frac{e^{-\beta \varepsilon_j}}{Z_B}}\langle\tilde i|\tilde U|\tilde j\rangle.
\label{kraussHeat-tilde}
\end{equation} 
Micro-reversibility implies that they satisfy  (see Appendix \ref{sec:appendix microrev})
\begin{equation}
\tilde M_{ji}=
\Theta_S\sqrt{e^{\beta(\varepsilon_i-\varepsilon_j)}}M_{ij}^\dag \Theta^\dag_S.
\label{Mmicroreversibility}
\end{equation} 
We can now relate the probability $p(\gamma)=|\langle b_m|M_{k=ij}|a_n\rangle|^2 p_i(n)$ for a trajectory $\gamma=\{n,k,m\}=\{n,i;m,j\}$ to the probability $\tilde p(\tilde\gamma)=|\langle\tilde a_n|\tilde M_{k=ji}|\tilde b_m\rangle|^2 \tilde p_f(m)$ of its time reversal $\tilde\gamma=\{\tilde m,\tilde k,\tilde n\}=\{\tilde m,\tilde j;\tilde n,\tilde i\}$. 
From Eq.~(\ref{Mmicroreversibility}), we have $\tilde p(\tilde \gamma)=e^{\beta(\varepsilon_i-\varepsilon_j)}|\langle a_n| M_{k=ij}^\dag| b_m\rangle|^2 \tilde p_f(m)$; therefore,
\begin{equation}
\frac{p(\gamma)}{\tilde p(\tilde \gamma)}=e^{-\beta(\varepsilon_i-\varepsilon_j)}\frac{p_i(n)}{\tilde p_f(m)}.
\label{FTgamma}
\end{equation}

If the initial state of the backward process $\tilde p_f(m)$ is the final state of the forward process, i.e., $\tilde p_f(m)=p_f(m)$, then we have $p(\gamma)=e^{\Delta_i s_\gamma}\tilde p(\tilde \gamma)$.
Using this equality, we now evaluate
 \[
 p(\Delta_is)=\sum_\gamma p(\gamma)\delta(\Delta_is-\Delta_is_\gamma)=e^{\Delta_i s}\sum_{\tilde\gamma}\tilde p(\tilde \gamma)\delta(\Delta_is+\Delta_is_{\tilde \gamma})=e^{\Delta_i s} \tilde p(-\Delta_i s),
 \]
where we have also used $\Delta_is_{\tilde \gamma}=-\Delta_is_\gamma$; see Eq.~(\ref{epgamma}). 

If the reversed process is identical to the forward process, $\tilde p(\Delta_i s)=p(\Delta_i s)$, then the fluctuation theorem for the entropy production, Eq.~(\ref{ep.fluct}), can be written just with the distribution of the forward process, $p(\Delta_i s)= e^{\Delta_i s} p(-\Delta_i s)$. This is the case in the systems that we consider if the driving is time symmetric, $\{H_S(t),V(t)\}=\{H_S(\tau-t),V(\tau-t)\}$ for $0\leq t<\tau$. A constant $H_S$ and (step) $V$ fulfill this condition. 
It is also necessary for the measured operator to be invariant under the time reversal transformation, guaranteeing a one-to-one correspondence between the forward and backward trajectories. See Appendix \ref{appendix: bkwequivalence}.

The detailed fluctuation theorem for the entropy production implies the integral fluctuation theorem $\langle e^{-\Delta_is}\rangle=1$. This in turn implies that if the average entropy production $\Delta_iS=\langle \Delta_is\rangle=0$ vanishes, then $\langle e^{-\Delta_is}\rangle=e^{\langle-\Delta_is\rangle}$, and due to the convexity of the exponential, this is possible only if $p(\Delta_is)=\delta(\Delta_is)$, i.e., the stochastic entropy production does not fluctuate. This was already noted when we discussed maps with equilibrium states and the fluctuation properties in these states. Conversely, it implies the opposite for {\it non-equilibrium} steady states, that is, fluctuations of the stochastic entropy production are necessary to have a positive average entropy production $\Delta_iS=\langle \Delta_is\rangle>0$.

\subsection{Fluctuations of work}

Now consider the case where the initial states of the forward and backward processes are canonical, $p_i(n)=e^{-\beta \epsilon_n}/Z_S$ and $\tilde p_f(m)=e^{-\beta \epsilon_m}/Z_S$; then, Eq.~(\ref{FTgamma}) 
provides $\frac{p(\gamma)}{\tilde p(\tilde \gamma)}=e^{-\beta(\varepsilon_i-\varepsilon_j)}e^{-\beta(\epsilon_n-\epsilon_m)}=e^{\beta w_\gamma}$. 
One can prove that if the reversed process is identical to the forward process (see Appendix \ref{appendix: bkwequivalence}), then the probability of
performing a work $w$ between the initial time with the system in the state $e^{-\beta H_S}/Z_S$ and an arbitrary time (possible after infinite time, when the system reaches the steady state) at which the energy of the system is measured satisfies the fluctuation relation
\begin{equation}
p(w)=p(-w)e^{\beta w}.
\label{WFT}
\end{equation}
Conversely, if the reversed process is not the same as the forward process, then the work fluctuation theorem reads $p(w)=\tilde p(-w)e^{\beta w}$.
For thermal maps, whose stationary state is the canonical thermal state, we observed that $p(w)=\delta(w)$; thus, Eq.~(\ref{WFT}) is trivially satisfied. For other maps, the canonical thermal state is not necessarily invariant. Thus, one can consider the evolution of the system initially prepared in the canonical thermal state toward its steady state and perform the two-point measurement of the system Hamiltonian to find that the work statistics follows Eq.~(\ref{WFT}). 
We later illustrate this equality in two interesting situations: a system that undergoes a cyclic process and a system with a NESS.

\subsection{Generalization to concatenated CPTP maps and applications}
\label{sec.appl}

In the previous section, we considered that the process $\rho_S\to\rho_S'$ is given by a single CPTP map $\rho_S'={\mathcal E}(\rho_S)=\sum_k M_k\rho_SM_k^\dag$ with a particular choice for the Kraus operators $M_k$ that allows a thermodynamic interpretation. We will show that the previous results of this section can be extended to concatenations of maps, providing a richer setup for studying thermodynamic processes.

\begin{figure}[H] 
\centering
  \includegraphics[width=0.4\textwidth]{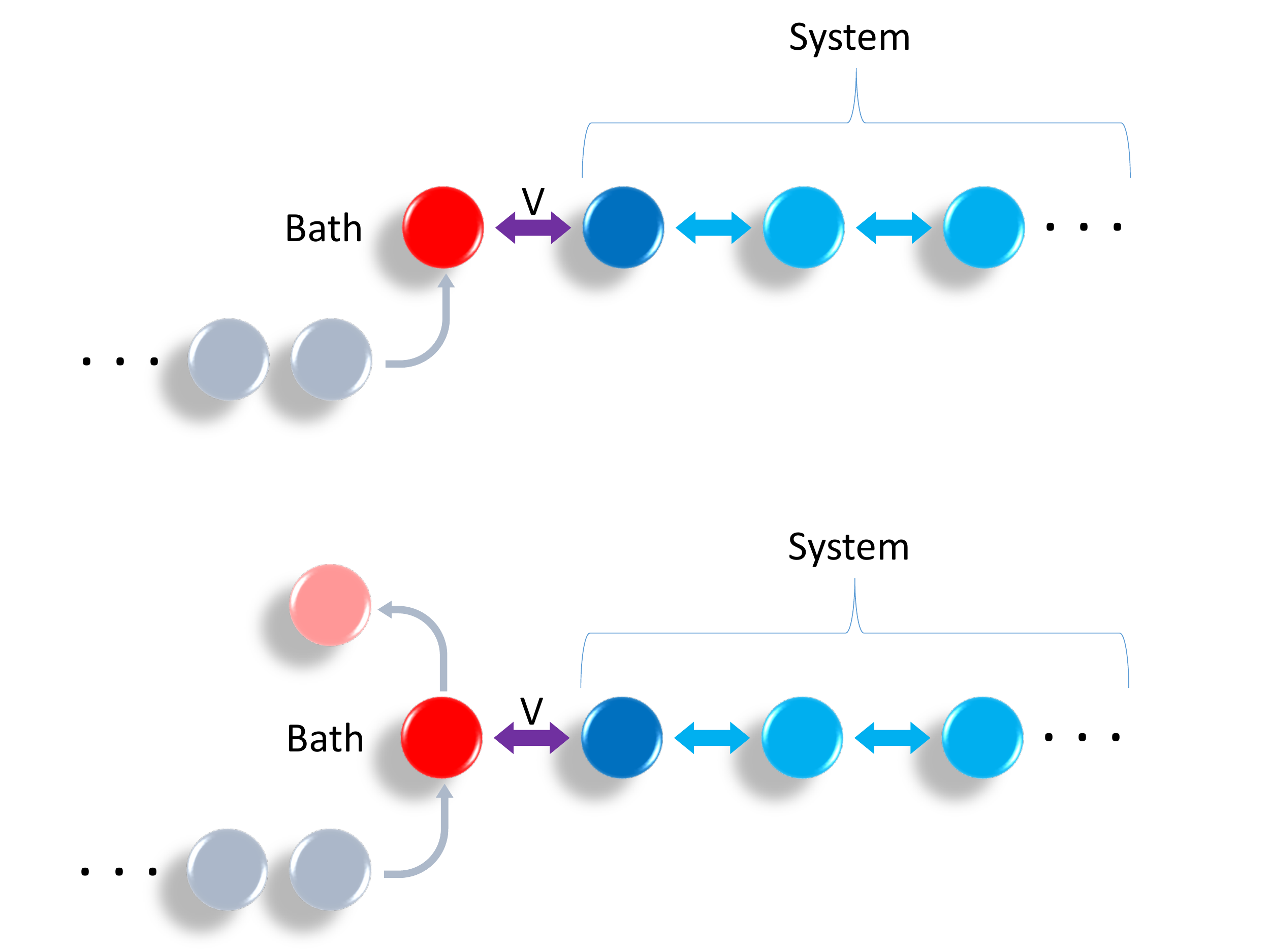}
    \caption{The figure depicts the first two interactions between the system and the copies of the bath.}
    \label{fig:esquema}
\end{figure}

One can concatenate CPTP maps 
acting over a system to describe a sequence of  evolutions of a system coupled to heat baths for given lapses of time. 
We generalize the concept of quantum trajectory to concatenations of $N$ maps $\mathcal{E}(\cdot)=\mathcal{E}^{(N)}\cdots\mathcal{E}^{(1)}(\cdot)$. 
Each 
\begin{equation}
\mathcal{E}^{(n)}(\cdot)=\sum_k M_k^{(n)}\cdot M_k^{(n)\dag}
\label{CPTP2}
\end{equation}
is a CPTP map, and for each, we measure a corresponding $k_n$ associated with the process $|i_n\rangle\to |j_n\rangle$ between eigenstates of the bath.
Note that with each map $\mathcal{E}^{(n)}$,  a new fresh bath is introduced to interact with the system. 
Fig.~\ref{fig:esquema} provides a scheme. 
As before, we consider the unitary evolution operator $U_n=e^{-i\tau_n(H_S+H_b^{n}+V^{n})}$, where $V^n$ represents the energy coupling between the system and the $n$th copy of the bath (with Hamiltonian $H_b^{n}$) in the time interval 
$[\sum_{l<n}\tau_l,\sum_{l\leq n}\tau_l]$.
We consider $V^{n}$ to be constant in this time interval and $V^{n}=0$ outside the interval. The Kraus operators are
\begin{equation}
M^n_{ij}=\sqrt{\frac{e^{-\beta \varepsilon_{i_n}}}{Z_b}}\langle j_n|U_n|i_n\rangle,
\label{kraussHeat2}
\end{equation}
where $\{\varepsilon_{i_n},|i_n\rangle\}$ is the spectrum of $H_b^n$ and $Z_b={\rm Tr} \,e^{-\beta H_b^n}$.
The stochastic heat flow from the bath to the system is given by minus the energy change of the baths $q_\gamma=\sum_n(\varepsilon_{i_n}-\varepsilon_{j_n})$. 
We perform a measurement of a system operator $A$ at the beginning and another $B$ at the end of the process. 
The trajectory is $\gamma=\{n,k_1,\ldots,k_N,m\}$, and its probability $p(\gamma)=p(m,k_1,\ldots,k_N|n)p_i(n)$ is
\begin{equation}
 p(\gamma)=|\langle b_m|M^{(N)}_{k_N}\cdots M^{(1)}_{k_1}|a_n\rangle|^2p_i(n)
\label{pgamma}
\end{equation}
or explicitly in terms of Eq.~(\ref{CPTP2}) and Eq.~(\ref{kraussHeat2}), the probability of
a trajectory $\gamma=\{a_n;i_1,j_1;\ldots;i_N,j_N;b_m\}$ is
\begin{equation}
p(\gamma)=|\langle b_m,j_1\cdots j_N|U_N\cdots U_1|i_1\cdots i_N,a_n\rangle|^2 \frac{e^{-\beta \sum_{n=1}^N\varepsilon_{i_n}}}{Z_b^N}p_i(n).
\label{psmN}
\end{equation}

With this, the detailed fluctuation theorem can be extended to concatenations of maps. If in the forward process the sequence ${\mathcal E^{(N)}}\cdots {\mathcal E^{(1)}}$ acts on an initial state, then the backward process is the reversed concatenation of the reversed maps, i.e., $\tilde{\mathcal E}^{(1)}\cdots \tilde{\mathcal E}^{(N)}$, and for a given trajectory, $\gamma=\{n,k_1\cdots k_N,m\}$, the corresponding backward trajectory is $\tilde\gamma=\{\tilde m,\tilde k_N\cdots \tilde k_1,\tilde n\}$.
The probability of the forward path is given by Eq.~(\ref{pgamma}), whereas for the backward path, the probability is
\begin{equation}
\tilde p(\tilde\gamma)=|\langle \tilde a_n|\tilde M^{(1)}_{k_1}\cdots \tilde M^{(N)}_{k_N}|\tilde b_m\rangle|^2 \tilde p_f(m)
\label{pgammatilde}
\end{equation}
Since every Kraus operator involved in Eq.~(\ref{pgammatilde}) satisfies Eq.~(\ref{Mmicroreversibility}), one obtains 
\begin{equation}
\frac{p(\gamma)}{\tilde p(\tilde \gamma)}=e^{-\beta q_\gamma}\frac{p_i(n)}{\tilde p_f(m)}
\label{FTgamma2}
\end{equation}
and as before, considering the initial state of the backward process $\tilde p_f(m)$ as the final state of the forward process, i.e., $\tilde p_f(m)=p_f(m)$, we have $p(\gamma)=e^{\Delta_i s_\gamma}\tilde p(\tilde \gamma)$. 
Finally, if the initial states of the forward and backward processes are canonical, $p_i(n)=e^{-\beta \epsilon_n}/Z_S$ and $\tilde p_f(m)=e^{-\beta \epsilon_m}/Z_S$, then Eq.~(\ref{FTgamma2}) 
provides $\frac{p(\gamma)}{\tilde p(\tilde \gamma)}=e^{-\beta q_\gamma}e^{-\beta(\epsilon_n-\epsilon_m)}=e^{\beta w_\gamma}$.

Note that for every iteration of the map, there is a certain amount of average work, heat and entropy production. These quantities are additive. This means that if we know the average work, heat and/or entropy production for two maps that are composed, then the average work, heat and/or entropy production for the total map (the composition) is the sum of the corresponding quantity for each map. 
For fluctuations, this separation is not possible. 

\subsection{A thermodynamic cycle}
\label{sec.cycle}
We can illustrate the advantage of considering concatenations of maps by studying a simple thermodynamic cycle. A system starts in the canonical thermal state $\omega_\beta(H_S)$, and then it is driven by a map such as the ones considered in Sec. \ref{traj.map.sec}. In such maps, work is performed on the system and leaves it in a non-equilibrium state $\rho_S'$ but with the same Hamiltonian $H_S$. Then, we assume that a thermal map brings the system back to the thermal state $\omega_\beta(H_S)$. 
For such a cycle, we have that 
\[
\Delta E=0=W_d+Q_d+Q_r,
\]
where $Q_d$ and $W_d$ respectively refer to the heat exchanged with the bath, Eq.~(\ref{Av.Heat}), and the work performed on the system, Eq.~(\ref{Av.Work}), during the driving process $\omega_\beta(H_S)\to \rho'_S$,
and $Q_r$ is the heat exchanged during the final relaxation process $\rho_S'\to \omega_\beta(H_S)$. Since this is achieved with a thermal map, one has $W_r=0$; thus, $Q_r={\rm Tr}[H_S(\omega_\beta(H_S)-\rho_S')]$.
One can verify that
\[
\Delta S=0=\Delta_iS_d+\Delta_iS_r+\beta(Q_d+Q_r),
\]
where $\Delta_iS_d=D(\rho'_{\rm tot}||\rho_S'\otimes\omega_\beta(H_B))$ and $\Delta_iS_r=D(\rho_S'||\omega_\beta(H_S))$ according to Eq.~(\ref{Av.Ent.Prod}) and Eq.~(\ref{epthermal}), respectively. 
It also follows that the total entropy production $\Delta_iS_d+\Delta_iS_r=\beta W_d$ is the dissipated work, as expected for an isothermal process starting and finishing in equilibrium. We also obtain $\Delta_iS_d=-D(\rho_S'||\omega_\beta(H_S))+\beta W_d$~\cite{DeffnerPRL}.

We now consider fluctuations. Note that because the initial and final density matrices are Gibbsian, work and entropy production fluctuations are identical,
\begin{equation} \label{ecc:thermo_cycle}
\ln \frac{p(\gamma)}{\tilde p (\tilde \gamma)} = \Delta_i s_\gamma = \Delta s_\gamma - \beta(q^d_\gamma +q^r_\gamma) = \beta \Delta \epsilon_\gamma -\beta( q^d_\gamma + q^r_\gamma) = \beta w_\gamma.
\end{equation}
In the last equation, $q_\gamma^{d,r}$ are the driving and relaxation stochastic heats, respectively. 

In general, the thermal state is achieved asymptotically 
when a system interacts with a large memoryless heat bath. This thermalization might also be achieved by concatenating (in theory) an infinite sequence of thermal maps, taking the limit $N\to \infty$ in the process described with Eqs.~(\ref{CPTP2} and \ref{kraussHeat2}). This composition is also a thermal map. In the first example presented in the next section, we show that when the system is small, this can be done much quicker in practice, with only a few maps.

\section{Examples}
\label{sec.aps}

\subsection{Single spin in a cycle}

\begin{figure}[H] 
\centering
  \includegraphics[width=0.4\textwidth]{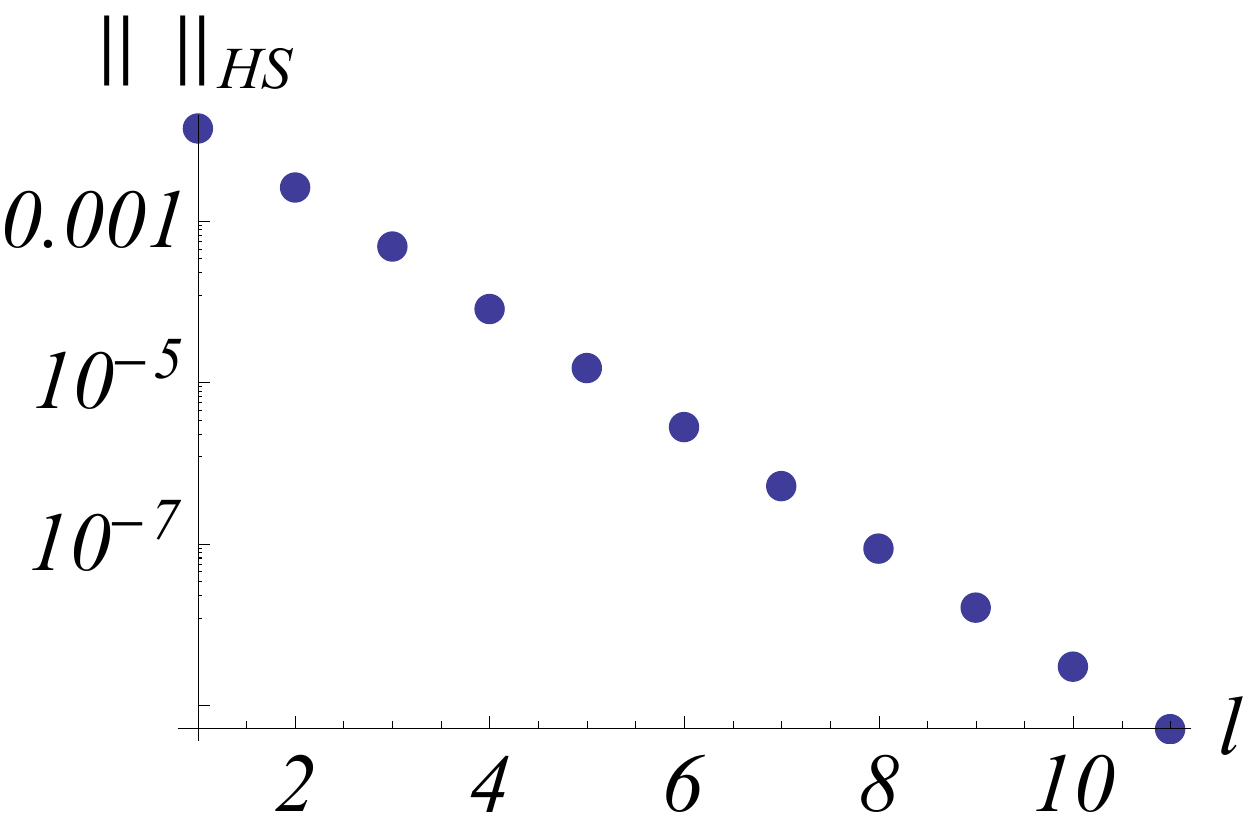}
  \hspace*{\fill}
    \includegraphics[width=0.4\textwidth]{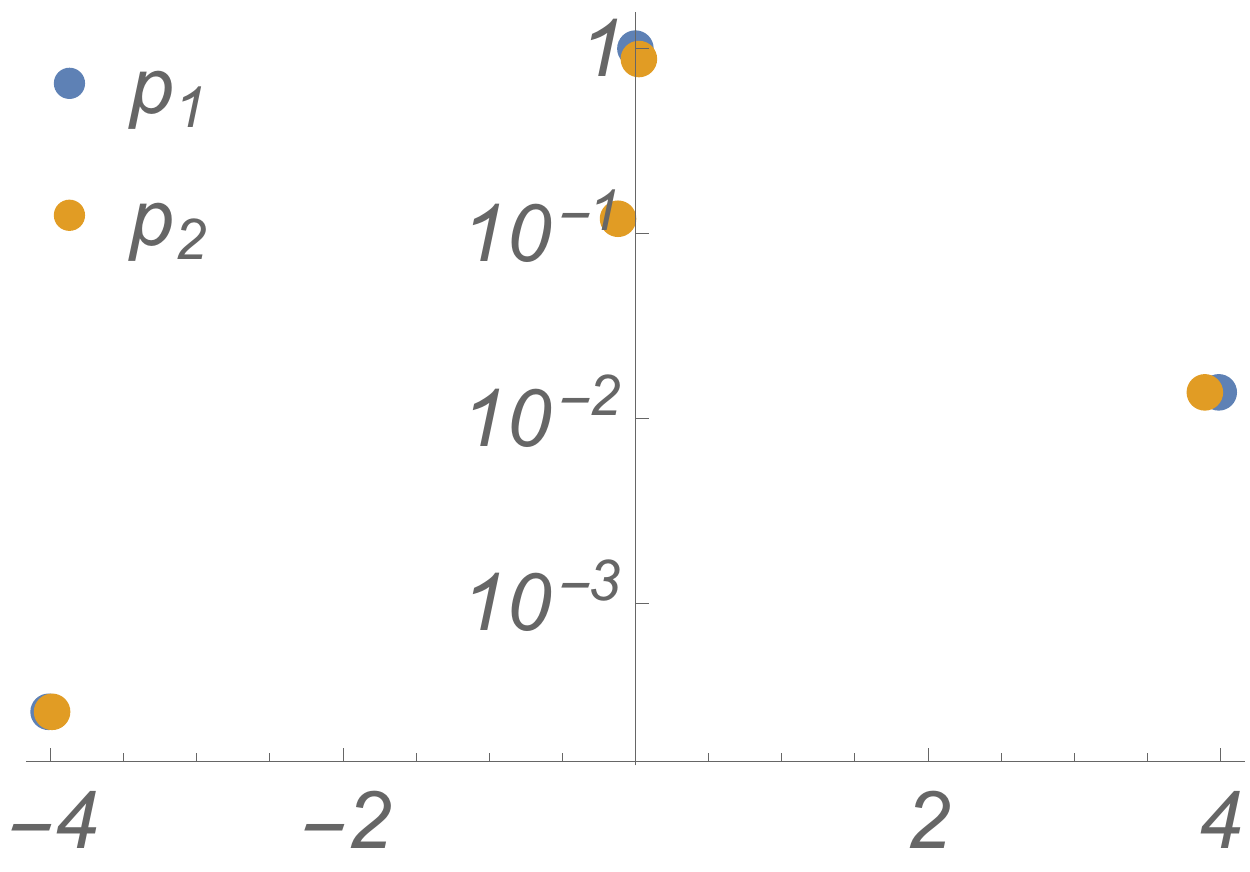}
    \caption{Left panel) Hilbert-Schmidt distance between the state of the system at each step of the concatenation and the thermal state. In this concatenation process, we take
    $\tau_1=1$, $V^1=(J_B + 0.3)\sigma^x_b\sigma^x_S + J_B \sigma^y_b\sigma^y_S$, $\tau_n=4, V^n=J_B \sigma^x_b\sigma^x_S + J_B \sigma^y_b\sigma^y_S$ for $n\geq 2$,
    and the other parameters are 
    $h=1$, $J_B=3$, and $\beta = 1.$ 
    Right panel) Work and entropy production distributions for the full cycle and the driving alone, $p_1 \equiv p_{\text{drive}}(\beta w) = p_{\text{cycle}}(\beta w)=p_{\text{cycle}}(\Delta_i s)\neq p_2 \equiv p_{\text{drive}}(\Delta_i s)$. The parameters of the plot are the same as in the left panel.}
    \label{fig:cycle}
\end{figure}

Consider a single spin with $H_S=(h/2)\sigma_S^z$ that interacts with thermal spins $\omega_\beta(H^n_b)$ with $H^n_b=(h/2)\sigma_b^z$. Here, $\sigma_{S,b}^{x,y,z}$ are the Pauli spin 1/2 operators of the system and bath. The interaction $V^1$ with the first thermal spin is such that $[H_S+H^1_b, V^1]\neq 0$; thus, if the spin system starts in equilibrium, $\rho_S = \omega_\beta(H_S)$, the interaction with the first thermal spin drives the system out of equilibrium to state ${\mathcal E}^{(1)}(\omega_\beta(H_S))$. In this process, some work is performed on the system and heat flows to the bath. Then, for the subsequent interactions, we take $V^n$ such that $[H_S+H^n_b, V^n]=0$ with $n\geq 2$;  thus, the corresponding $\{{\mathcal E}^{(n)}\}_{n\geq 2}$ are thermal maps that will bring the system back to the thermal $\omega_\beta(H_S)$ state without performing or extracting work.
This constitutes a thermodynamic cycle for the system with a unitary evolution for the total system of spin plus baths. Consequently, the micro-reversibility principle is valid, and Eq.~(\ref{ecc:thermo_cycle}) is fulfilled (see the next example for details on the time reversal operator).
In the left panel of Fig.~\ref{fig:cycle}, we plot the Hilbert-Schmidt distance \footnote{Hilbert-Schmidt distance is defined as $||\sigma-\rho||_\text{HS} \equiv \Tr[(\sigma-\rho)^\dag (\sigma-\rho)]$.} between the state of the system at each step of the concatenation and the thermal state,
and we indeed observe that as $N$ increases, thermalization becomes 
more effective. In the right panel, we show  work and entropy production probability distributions for the cycle considering that at $N=7$, thermalization has been achieved ($||\rho_S'-\omega_\beta(H_S)||_\text{HS}<10^{-5}$). We show in Appendix \ref{appe.thermal maps dont contribute} that thermal maps do not contribute to work fluctuations; thus, we plot work fluctuations of the driving part alone, i.e., ${\mathcal E}^{(1)}$, as well as of the full cycle, $p_1 \equiv p_\text{cycle}(\beta w)=p_\text{drive}(\beta w) = p_\text{cycle}(\Delta_i s)$. Note that this does not apply for entropy production, $p_2 \equiv p_\text{drive}(\Delta_i s) \neq p_1$.

\subsection{Spin 1/2 chains}

Let us consider a one-dimensional spin $1/2$ chain $\vec\sigma_1\cdots\vec\sigma_N$ with Hamiltonian $H_S$ interacting through the first site with 
a spin $1/2$ particle with Hamiltonian $H_b=\frac{h}{2}\sigma_b^z$ in a thermal state, where the interaction is given by 
\begin{equation}
V=J_B(\sigma_b^x\sigma_1^x+\sigma_b^y\sigma_1^y).
\label{lambda}
\end{equation}
Here, the Pauli operators $\sigma_b^{x,y,z}$ belong to the single bath spin, and $\sigma_1^{x,y,z}$ belong to the first spin of the chain. 
The unitary evolution for the system plus bath is given by the operator $U=e^{-i\tau(H_S+H_b+V)}$. Micro-reversibility holds if an anti-unitary operator $\Theta$ exists such that
$\Theta \tilde U\Theta^\dag =U^\dag$. For the spin systems that we consider below, $\Theta=i\sigma_b^x i \sigma_b^y\Pi_{i=1}^{N}(i\sigma^x_n i \sigma^y_n) K$ is a time reversal symmetry operator, where $K$ performs the complex conjugation. Note that when a system involves a magnetic field, one generally needs to invert the direction of the magnetic field; thus, the detailed fluctuation theorem  in that case will relate the fluctuations of two different systems. However, for the time reversal operator that we consider here, that is not the case \cite{Haake}. The factor $i\sigma^x$ rotates the system $180\degree$ in the $x$ direction, leaving the $\sigma^z$ terms invariant (see the Hamiltonians below).
By replacing $H_b$ in Eq.~(\ref{kraussHeat}), one obtains four Kraus operators $M_{ij}$ with $i,j=\pm$, which correspond to transitions in the bath when $i\neq j$.

\begin{figure}[H] 
\centering
  \includegraphics[width=0.4\textwidth]{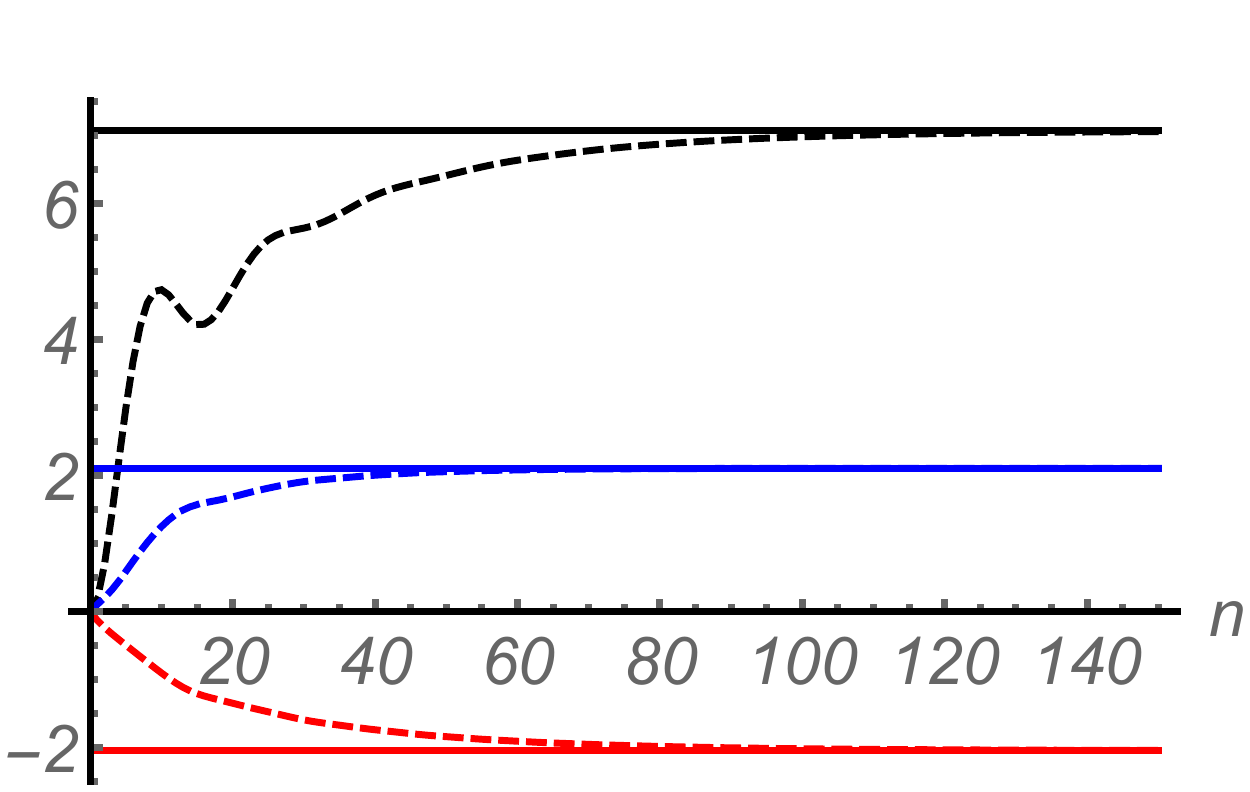}
  \hspace*{\fill}
    \includegraphics[width=0.4\textwidth]{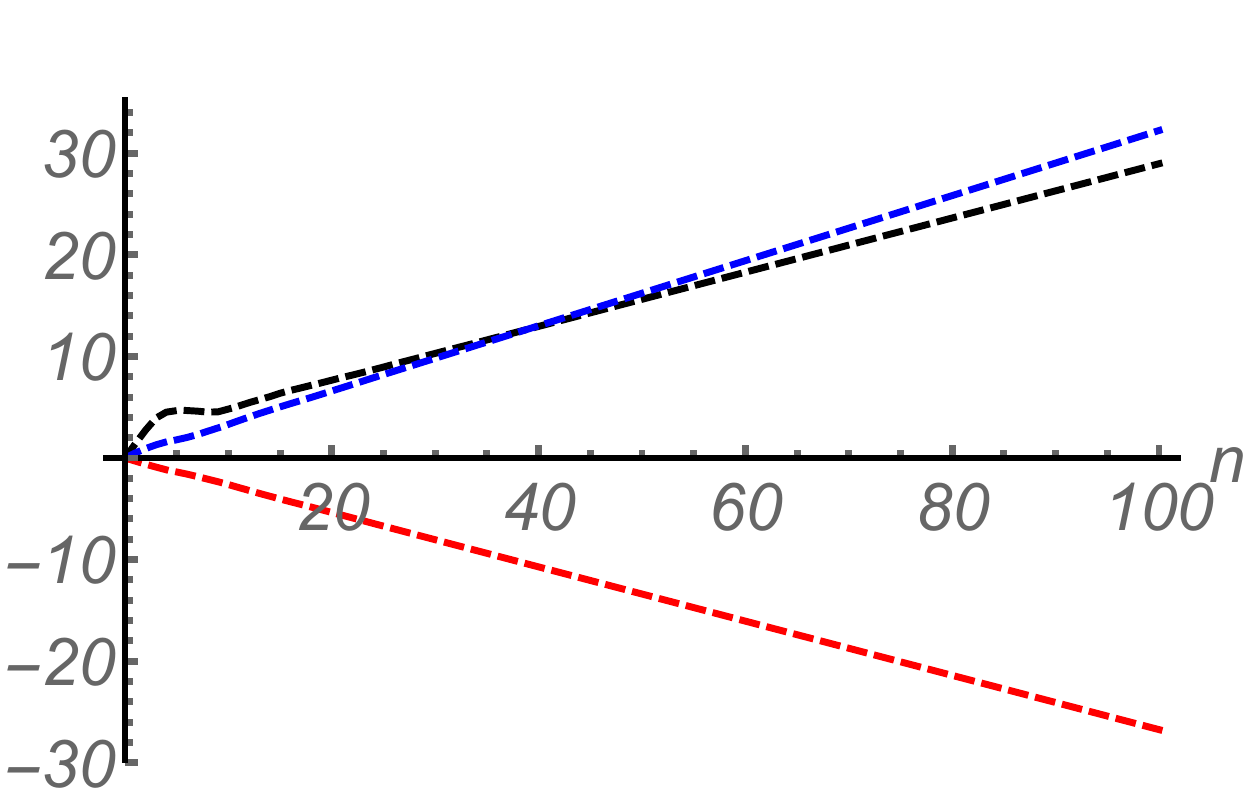}
    \caption{Left panel) For the homogeneous XX spin 1/2 chain with three sites, cumulated average work (dashed-black), heat (dashed-red) and entropy production (dashed-blue) as a function of the iteration number. The corresponding straight lines are the theoretically computed asymptotic values.  Right panel) The same quantities for the homogeneous XY spin 1/2 chain with three sites.
In both cases, the initial density matrix is the Gibbs state $\omega_\beta(H_S)$, which is a non-equilibrium state. The parameters for the plots are $h=2$, $J_i^x=J_B=3$, $\beta = 1.2$ and $\tau =1$, and $J_i^y=2$ in the right panel.}
    \label{fig:Thermo}
\end{figure}

\begin{figure}[H] 
\centering
  \includegraphics[width=0.4\textwidth]{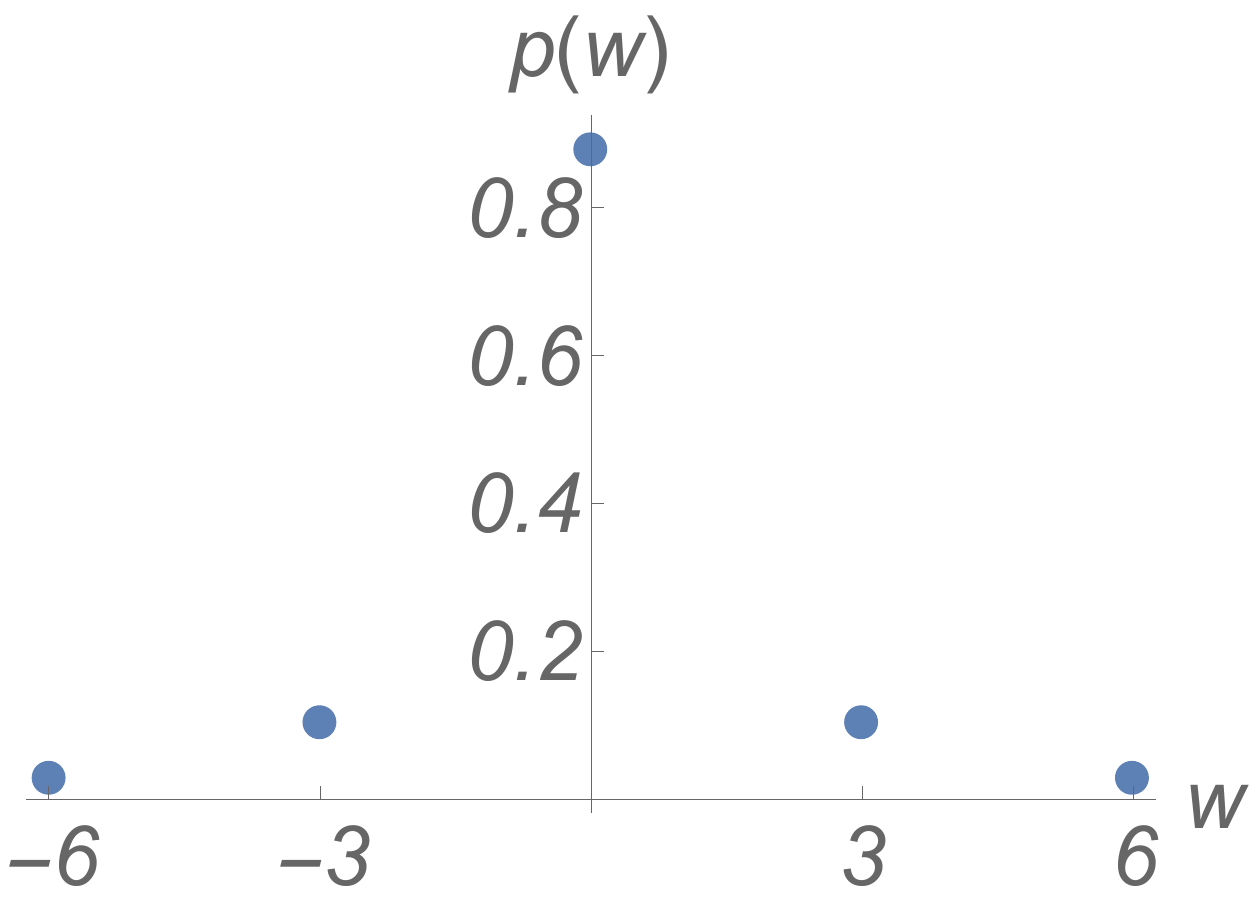}
  \hspace*{\fill}
    \includegraphics[width=0.4\textwidth]{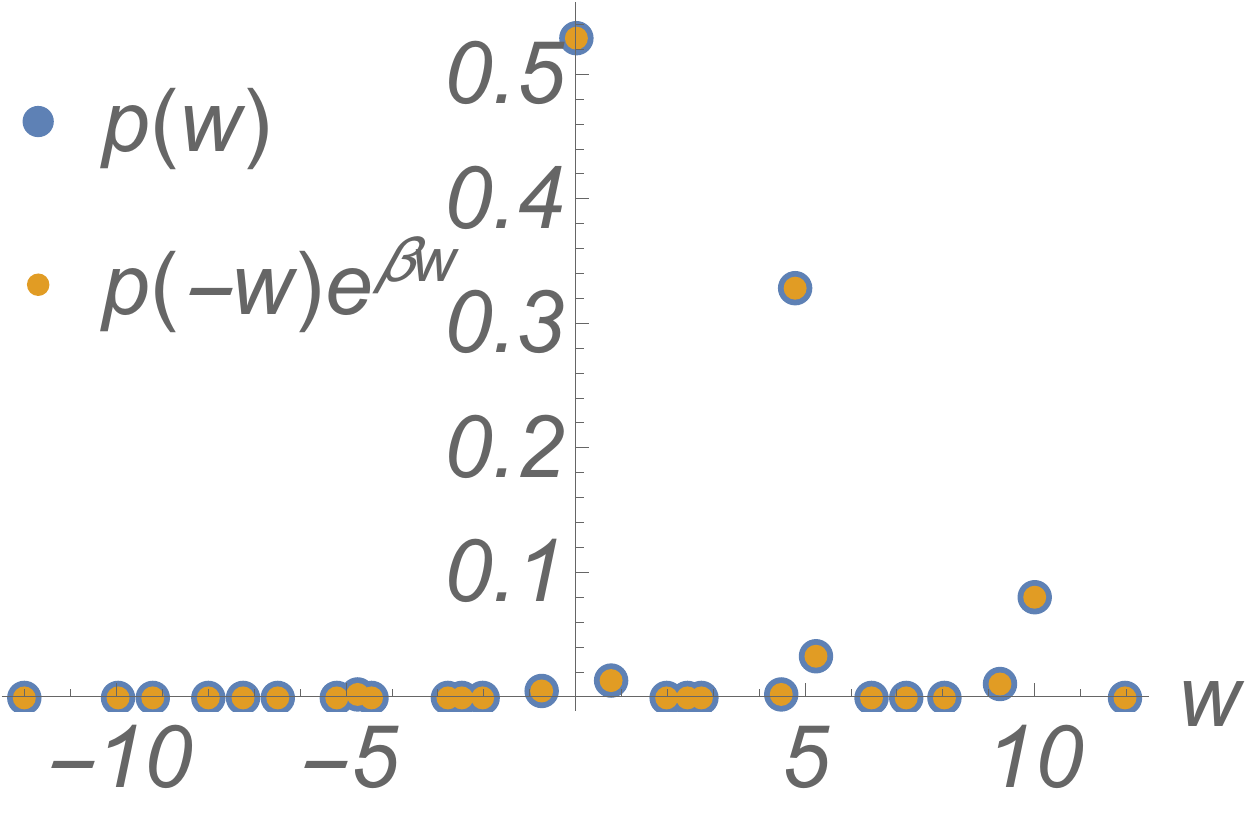}
    \caption{Left panel) Work distribution for the XX spin 1/2 chain in the equilibrium state. 
    Right panel) Work fluctuation relation Eq.~(\ref{WFT}) for the XY spin 1/2 chain, starting from the Gibbs state in the forward and backward processes. In both plots, we have used a chain of two sites and one map. The parameters for the plots are $h=2$, $J_i^x=J_B=3$, $\beta = 1.2$ and $\tau =1$, and $J_i^y=2$ in the right panel.}
    \label{fig:wFTXY}
\end{figure}

Now let the system be an XX spin $1/2$ chain with Hamiltonian
\[
H_{XX}=\frac{h}{2}\sum_{i=1}^M\sigma_i^z - \sum_{i=1}^{M-1} J_i^x(\sigma_{i}^x\sigma_{i+1}^x+\sigma_{i}^y\sigma_{i+1}^y).
\]  
The total magnetization is a conserved quantity for the XX spin chain, i.e.,  $[H_{XX},H_0]=0$, where $H_0=\frac{h}{2}\sum_{i=1}^M \sigma_i^z$ is the non-interacting part of the Hamiltonian. Considering the Hamiltonian of the bath $H_b=\frac{h}{2}\sigma_b^z$ and the interaction between the chain and the bath $V$ in Eq.~(\ref{lambda}), one finds that the unitary evolution $U=e^{-i\tau(H_{XX}+H_b+V)}$ satisfies $[U,H_0+H_b]=0$; thus, $\omega_\beta(H_0)$ is an equilibrium state of the map ${\mathcal E}$. 
In this example, the Hamiltonian is of the form $H_S=H_0+H_I$, where $[H_0,H_I]=0$. The stochastic thermodynamics for maps with equilibrium was discussed in section \ref{stoch.term.sec}. 
We can observe that by iterating the map as discussed in section \ref{sec.appl}, an initial state $\rho_S(0)$ converges to the equilibrium state $\omega_\beta(H_0)$. This is illustrated in Fig.~\ref{fig:Thermo}. The (cumulated) average work performed on that process is given by Eq.(\ref{Av.Work.Eq}), and because  ${\rm Tr}_S[H_I\omega_\beta(H_0)]=0$ in this example, it is simply given by $W=-{\rm Tr}_S[H_I\rho_S(0)]$,
the (cumulated) heat is $Q={\rm Tr}_S[H_0(\omega_\beta(H_0)-\rho_S(0))]$, and the (cumulated) entropy production is $\Delta_iS=D(\rho_S(0)||\omega_\beta(H_0))$. The asymptotic values are indicated in Fig.~\ref{fig:Thermo}, left panel.
Regarding the fluctuating properties, we showed in \ref{stoch.term.sec} that in the equilibrium state $\omega_\beta(H_0)$, the entropy production does not fluctuate, but work may fluctuate. This is illustrated in Fig.~\ref{fig:wFTXY}, left panel.

Let us now consider the XY spin $1/2$ chain with Hamiltonian
\[
H_{XY}=\frac{h}{2}\sum_{i=1}^M\sigma_i^z - \sum_{i=1}^{M-1} (J_i^x \sigma_{i}^x\sigma_{i+1}^x+J_i^y \sigma_{i}^y\sigma_{i+1}^y)
\]  
coupled to the bath with Hamiltonian $H_b$ through the same coupling $V$. For this system, the invariant state $\rho_{XY}$ of the map ${\mathcal E}$ is not related to any operator $H_0$ such that $[U,H_0+H_b]=0$.
Indeed, the state  $\rho_{XY}\otimes\omega_\beta(H_b)$ is not invariant under the unitary evolution, and this already indicates that the steady state $\rho_{XY}$ is a NESS. We can observe that by iterating the map for the XY chain, the steady state is reached, where a constant amount of work is being performed each time that the map is applied and the same for the heat and entropy production. This is illustrated by the constant slopes in the cumulated thermodynamic quantities in Fig.~\ref{fig:Thermo}, right panel. 
These slopes can be computed, but not in terms of the system properties. They are global quantities. For instance, the slope for the entropy production is $D(U\rho_{XY}\otimes\omega_\beta(H_b)U^\dag||\rho_{XY}\otimes\omega_\beta(H_b))$. In this case, the map on the XY spin chain is a map with NESS.

Let us now consider some fluctuation properties.
In the left panel of Fig.~\ref{fig:wFTXY}, we plot the work distribution for the XX spin $1/2$ chain in the equilibrium state,
where $p(\Delta_is)= \delta(\Delta_is)$, but as noted previously, the work fluctuates, i.e., $p(w)\neq \delta(w)$, even though $\langle w\rangle=0$. 
The work fluctuation relation, Eq.~(\ref{WFT}), is illustrated in the right panel of Fig.~\ref{fig:wFTXY} for the XY spin chain. 

To conclude this section, we note that for the XX case, where the map has equilibrium, and for the XY case, where the map has a NESS, 
the agent performs (or extracts) work. This work is easily understood as the cost of implementing the concatenation, i.e., due to the (periodic) time dependence of the coupling between the system and the bath.

\section{Consequences for Lindblad Dynamics} \label{sec: Lindblad}

In this section, we present the consequences of our previous results for systems described with some Lindblad equations. 
We will only consider the averaged thermodynamic quantities discussed previously and omit the fluctuations. 

As discussed in \cite{Attal, Karevski}, when the interaction strength between the system and the bath is scaled as $V=v/\sqrt{\tau}$ in the scheme discussed above, where $\tau$ is the time duration of the coupling between the system and the bath in every iteration, a Lindblad equation describes the dynamics in the continuous time limit $\tau \to 0$. In \cite{Barra}, the limit was considered  for the dynamics and the thermodynamic
quantities simultaneously. 

Recalling that the maps that we consider are given by Eq.~(\ref{CPTP}), we first note that by scaling the coupling as $V=v/\sqrt{\tau}$, the map $\mathcal E$ can be written up to order $\tau$ as
\begin{equation} \label{ecc: lindblad: repeated infinitesimal}
\mathcal E^{\tau}(\rho_S) = \rho_S - i\tau [H_S, \rho_S] +  \tau \sum_{ij} \frac{e^{-\beta \varepsilon_i}}{Z_B} \left( L_{ij} \rho_S L_{ij}^\dag - \frac{1}{2}\{ L_{ij}^\dag L_{ij},\rho_S\} \right) ,
\end{equation}
where $L_{ij}= \bra{j}v\ket{i}$ ($v$ is Hermitian) and $H_B\ket{i} = \varepsilon_i\ket{i}$. The double sum in the last term of Eq.~(\ref{ecc: lindblad: repeated infinitesimal})
is split into the diagonal part, $i=j$, and the non-diagonal part, $i\neq j$, which in turn can be converted into a sum $\sum_{i<j}$ by combining the $ij$ and the $ji$ terms. 
Note that if we consider the labels $\{ij\}$ with $i\le j$ as one label,  $r=ij$, then the last term of Eq.~(\ref{ecc: lindblad: repeated infinitesimal}) has the form $\tau\mathcal D(\rho_S)$, where
\begin{equation} \label{ecc: dissipator theorem}
\mathcal D(\cdot) = \sum_r \gamma_r \left[L_r \cdot L_r^\dag - \frac{1}{2}\{L_r^\dag L_r, \cdot \} + \omega_r^{-1} \left(L_r^\dag \cdot L_r - \frac{1}{2}\{L_r L_r^\dag, \cdot\}\right) \right]
\end{equation}
is a Lindblad dissipator with Lindblad operators $L_r=L_{ij}$ and coefficients $\omega_r= e^{\beta(\varepsilon_j-\varepsilon_i)}$ and 
$\gamma_r = e^{-\beta \varepsilon_i}/Z_B>0$ when $i< j$ and $\gamma_r = e^{-\beta \varepsilon_i}/2Z_B>0$ when $i=j$. Considering $\partial_t \rho_S=\lim_{\tau\to 0}\frac{\mathcal E^{\tau}(\rho_S)-\rho_S}{\tau}$,
we have the Lindblad evolution equation
\begin{equation}
\partial_t \rho_S=- i [H_S, \rho_S] +\mathcal D(\rho_S).
\label{LindEq}
\end{equation}
Note that in the general case, this Lindblad equation does not fulfill the quantum detailed balance condition~\cite{Alicki76} because there is no \textit{a priori} relation between the Lindblad operators $L_r$ and the system Hamiltonian $H_S$ (see \textit{Theorem 3} in~\cite{Alicki76}). 

If we simultaneously consider the limit for the thermodynamic quantities in Eqs.~(\ref{Av.Heat},~\ref{Av.Work}, and~\ref{Av.Ent.Prod}), then we obtain
\begin{equation}
\dot Q =-\Tr_S\left[\sum\limits_{ij} \frac{e^{-\beta\varepsilon_i}}{Z_B} \left(\varepsilon_j L_{ij} \rho_S L_{ij}^\dag - \frac{\varepsilon_i}{2}\{ L_{ij}^\dag L_{ij},\rho_S\} \right) \right], 
\label{newQ}
\end{equation}
\begin{align}
\dot W &= \Tr_S[H_S \mathcal D(\rho_S)] -\dot Q, \\
\dot S_i &= \Tr_S[\mathcal D(\rho_S) \ln \rho_S] - \beta \dot Q,
\end{align}
which correspond to the heat flux, the power performed on (or by) the system and the rate of entropy production for the process described by the Lindblad dynamics.

Remarkably, if $\mathcal E$ is a map with equilibrium, then the corresponding Lindblad equation $\partial_t \rho_S=\lim_{\tau\to 0}\frac{\mathcal E^{\tau}(\rho_S)-\rho_S}{\tau}$ has detailed balance. To prove this, one notes that the equilibrium condition implies $[H_0,v] = [v,H_B]$,
and therefore, 
\begin{align}
[H_0,L_{r}] &= (\varepsilon_i-\varepsilon_j)L_{r} \label{ecc:L_r eigenoperator}\\
[H_0,L_{r}^\dag] &= -(\varepsilon_i-\varepsilon_j)L_{r}^\dag. \label{ecc:L_rdag eigenoperator}
\end{align} 
These conditions, together with $[H_0,H_S]=0$, imply that the Lindblad equation, Eq.~(\ref{LindEq}), with dissipator, Eq.~(\ref{ecc: dissipator theorem}), has quantum detailed balance with respect to 
$\pi=e^{-\beta H_0}/Z_0$ \footnote{ It can be shown that the quantum detailed balance
condition [$\mathcal L_a(\pi)=0$ and $\mathcal L_s(A\pi)=\mathcal L_s^*(A)\pi$ for all observables $A$, where $\mathcal L_{(a)s}$ is the (anti-)symmetric part of $-i[H_S,\cdot]+\mathcal D(\cdot)$, see~\cite{Alicki76} for further details] can be generalized for equilibrium states $\pi=e^{-\beta H_0}/Z_0$ for the Lindblad Eq.(\ref{LindEq}) with dissipator Eq.(\ref{ecc: dissipator theorem}) when the following holds: $[H_0, L_r] = -\frac{1}{\beta} \ln \omega_r L_r $, $[H_0, L_r^\dag] = \frac{1}{\beta}\ln \omega_r L_r^\dag$ and $[H_0,H_S]=0.$ (In the particular case that $H_0=H_S$, the detailed balance relation is standard.)}.

If our Lindblad equation of interest has detailed balance with respect to a positive invariant equilibrium state, i.e., $\pi = \omega_\beta(H_0)$, then the thermodynamic quantities for CPTP maps, which are simplified due to an equilibrium, can be written in the following way in the Lindblad limit without requiring explicit knowledge of the particular form of $H_B$ and the coupling $V$~\footnote{These follow directly from Eqs.~(\ref{Av.Work.Eq},~\ref{Eq.prop}). One can also start from Eq.(\ref{newQ}), evaluate the trace in the eigenbasis of $H_S$ and use the equalities $(\epsilon_n^0-\epsilon_m^0)\bra{\epsilon_n}L_{ij}\ket{\epsilon_m}=(\varepsilon_i-\varepsilon_j)\bra{\epsilon_n}L_{ij}\ket{\epsilon_m}$ and $[H_0,L_{ij}^\dag L_{ij}]=0$, which are two consequences of Eqs.(\ref{ecc:L_r eigenoperator}-\ref{ecc:L_rdag eigenoperator}). }:
\begin{align}
\dot Q(t) &= \Tr[H_0 \mathcal D (\rho_S(t))] \label{ecc: simplified heat lindblad lim}\\
\dot W(t) &= \Tr[(H_S-H_0) \mathcal D(\rho_S(t))] \label{ecc: simplified work lindblad lim}\\
\dot{S}_i(t) &= - \Tr[\mathcal D (\rho_S(t)) \ln \rho_S(t)] - \beta \dot Q. \label{ecc: simplified ent. prod. lindblad lim}
\end{align}

Note that if $H_0=H_S$ and thus the equilibrium state for which detailed balance is valid is Gibbsian, $\pi = \omega_\beta(H_S)$, then $\dot W = 0$ according to Eq.~(\ref{ecc: simplified work lindblad lim}), and Eqs.~(\ref{ecc: simplified heat lindblad lim}) and (\ref{ecc: simplified ent. prod. lindblad lim}) provide the standard definitions used for the heat and entropy production in the weak-coupling regime \cite{BreuerBook}.

As an example, consider the spin 1/2 chains of the last section with the same bath and scaling of the coupling Eq.~(\ref{lambda}) as $J_B=\sqrt{\lambda/\tau}$. In the limit discussed above, one obtains the same Lindblad equation, other than the unitary part, for the $H_S=H_{XX}$ or $H_S=H_{XY}$ spin chains, which is given by

\begin{equation} \label{ecc: lindblad: lindblad XX or XY}
\partial_t \rho_S = -i[H_S,\rho_S] + \gamma_+\left[ \sigma_1^+ \rho_S \sigma_1^- - \frac{1}{2}\{\sigma_1^- \sigma_1^+,\rho_S\} + e^{\beta h} \left( \sigma_1^- \rho_S \sigma_1^+ - \frac{1}{2}\{\sigma_1^+ \sigma_1^-,\rho_S\} \right) \right],
\end{equation}
where $\gamma_+ = 2\lambda(1-\tanh(\beta h/2))$ and $\sigma_1^\pm=\sigma_1^x \pm i \sigma_1^y$. 
 Analogously to the case of the maps, the Lindblad equation for the XX chain has an equilibrium steady state $\pi=e^{-\beta H_0}/Z_0$ with $H_0 = (h/2)\sum_i \sigma_i^z$. Because $[H_0, \sigma_1^\pm] = \pm h\sigma_1^\pm$, Eqs.~(\ref{ecc:L_r eigenoperator}) and (\ref{ecc:L_rdag eigenoperator}) are verified, and since we also have that $[H_{XX},H_0]=0$, the Lindblad equation for the XX chain has quantum detailed balance with respect to the state $\pi=e^{-\beta H_0}/Z_0$.
An initial condition $\rho_S(0)$ will relax to the equilibrium state $\pi$ according to this Lindblad equation, and even though it is time independent, work has to be performed or extracted in this relaxation process according to Eq.~(\ref{ecc: simplified work lindblad lim}). 
In the case of the XY chain, one can find the steady state numerically and verify that it is a NESS, i.e., quantum detailed balance is not satisfied and work is continuously being done and dissipated as heat in the steady state. The work that accompanies the process is impossible to obtain from the sole knowledge of the Lindblad equation. We can physically understand this work
considering that this Lindblad dynamics arises due to the active role of an agent that continuously refreshes the bath that interacts with the system in such a way that he or she imposes
the form of the dissipator (the Lindblad operators $L_r$) that act on the system.

\section{Conclusion}
\label{secCONC}

We have studied the stochastic thermodynamics of CPTP quantum maps, specifically discussing the properties of maps with equilibrium. Thermal maps are a very important type of map with equilibrium because they represent the passive effect of a heat bath on a system, whereas non-thermal maps with equilibrium require the active intervention of an agent, manifested by the work required to perform the process represented by ${\mathcal E}$ even when the system Hamiltonian is non-driven. 

Non-thermal maps with equilibrium are, in a sense, between thermal maps and maps with non-equilibrium steady states. 
They share thermal maps' simplicity of relaxing to the equilibrium state, and at the same time, they allow us to study energy exchanges
between the system,  bath and experimenter in a non-trivial case.  
The results that we have illustrated here 
are not restricted to quantum systems, but in quantum systems, the manipulation of the interaction with the environment is much better controlled; thus, the possibility of transforming the state of a system using {\it active} interactions $V(t)$ with the bath, as illustrated with the XX spin 1/2 chain, is already in the tool box of the experimentalist~\cite{ion-trap}.  An implementation of the XX chain with trapped ions is possible, and it can be a candidate to investigate further aspects of quantum thermodynamics. The properties that we have derived for maps with equilibrium, namely, the local (in terms of system's operators) thermodynamic quantities and their fluctuation simplifications, could also be experimentally verified in NMR experiments \cite{NMRexp}.

Stochastic thermodynamics for classical systems with Hamiltonian or stochastic dynamics (Langevin or discrete master equation) is usually formulated for systems with the properties that we associate with thermal operations:  in the absence of driving, the system thermalizes to the corresponding Gibbs state, and the stochastic work is zero in non-driven systems. 
For instance, in classical stochastic thermodynamics, as well as for thermal maps, a system in equilibrium presents no entropy production fluctuation, which also implies no work fluctuations. In the quantum case, one can consider these fluctuations simultaneously if the equilibrium state commutes with the Hamiltonian of the system.  We have observed that in the XX spin 1/2 chain that we have studied, where the dynamics is controlled by a non-thermal map with equilibrium and where the equilibrium state commutes with the Hamiltonian, it is possible to have a fluctuating work and no entropy production fluctuations. 

Meanwhile, open systems in contact with a single heat bath, whose dynamics is represented by a map with NESS, present a complexity similar to the one of open systems passively coupled to two heat baths~\cite{pillet}. A few properties for this case were also studied, such as the statistics of work for systems starting in the Gibbs $\omega_\beta(H_S)$ state, Eq.~(\ref{WFT}). 

We showed that the results for a single map are also valid for concatenations of them. 
A process corresponding to a concatenation of maps is performed with some work cost (or gain) due to the switching on and off of the coupling between the system and the new and fresh copy of the bath, even for time independent $H_S$ and $V$.
When we iterate CPTP maps with equilibrium and take the continuous limit of Sec.~\ref{sec: Lindblad}, the resulting Lindblad equation has an equilibrium steady state and satisfies detailed balance respect to it. For CPTP maps with NESS  the resulting Lindblad equation has a NESS.
Remarkably, the thermodynamic quantities in the continuous limit can be written locally (in terms of the system's operators) when the map used has equilibrium. Normally, one cannot say much about the thermodynamics of an \textit{a priori} given Lindblad equation unless it satisfies the detailed balance condition with respect to the Gibbs state. 
An important conclusion from our analysis is that even though the given Lindblad equation may be time independent, i.e., no time-dependent driving in the system Hamiltonian and time-independent dissipative parts, the process described by the Lindblad equation is conducted by performing (or extracting) work, which we interpret as the work cost (or gain) of coupling the system to the bath with the given Lindblad operators. Lindblad operators arising from thermal maps have zero work cost but a Lindblad dynamics describing relaxation toward a non-Gibbsian equilibrium state occurs with a work contribution; see Eq.~(\ref{ecc: simplified work lindblad lim}). 
Our analysis (see also~\cite{Barra}) could provide a procedure to derive a consistent thermodynamics for a broad class of Lindblad evolutions.

\section*{Acknowledgments}

F.B. gratefully acknowledges the financial support of FONDECYT grant 1151390. C.L. gratefully acknowledges the financial support of CONICYT through Becas Magister Nacional 2016, Contract No. 22161809.

\section*{Appendices}
\begin{appendices}
\section{Proof of Eq.~(\ref{Mmicroreversibility})} \label{sec:appendix microrev}

Eq.~(\ref{Mmicroreversibility}) is an equality between operators in the system Hilbert space ${\mathcal H}_S$, which we can write as
\[
\bra{\tilde i}\tilde U\ket{\tilde j}=\Theta_S \bra{j} U\ket{i}^\dag \Theta_S^\dag,
\]
where $\tilde U=\Theta U^\dag\Theta^\dag$ and $\ket{\tilde i}=\Theta_B \ket{i}$. 
We prove it by showing that the matrix elements in an arbitrary basis are the same. Consider the operator at the left-hand side, and evaluate its matrix elements with states $\ket{\tilde a}=\Theta_S\ket{a}$ and $\ket{\tilde b}=\Theta_S\ket{b}$, i.e., 
\[
\left(\ket{\tilde a},\bra{\tilde i}\tilde U\ket{\tilde j}\ket{\tilde b}\right)_{\mathcal{H}_S}=\left(\ket{\tilde a\tilde i},\tilde U\ket{\tilde b\tilde j}\right)_{\mathcal{H}_{\rm tot}}=
\left(\Theta\ket{ a i},\Theta U^\dag\ket{ bj}\right)_{\mathcal{H}_{\rm tot}}=\left(U^\dag\ket{bj},\ket{ a i}\right)_{\mathcal{H}_{\rm tot}}=\bra{bj}U\ket{ai}.
\]
Let us now evaluate the same element for the operator on the right-hand side
\[
\left(\ket{\tilde a},\Theta_S \bra{j} U\ket{i}^\dag \Theta_S ^\dag \ket{\tilde b}\right)_{\mathcal{H}_S}=\left(\Theta_S\ket{ a},\Theta_S \bra{j} U\ket{i}^\dag \ket{ b}\right)_{\mathcal{H}_S}
=\left( \bra{j} U\ket{i}^\dag \ket{ b},\ket{ a}\right)_{\mathcal{H}_S}=\left(  \ket{ b},\bra{j} U\ket{i}\ket{ a}\right)_{\mathcal{H}_S}=\bra{bj}U\ket{ai},
\]
and therefore, we have the equality. Note that we have used the property $\left(\Theta \phi,\Theta\psi\right)=\left( \psi,\phi\right)$ that the anti-unitary operators $\Theta$ and $\Theta_S$ satisfy in the corresponding Hilbert space. 

\section{ $p(w) = \tilde p(w)$ if  the Hamiltonian $H_\text{tot}$ is invariant under time reversal.}
\label{appendix: bkwequivalence}

For the spin systems that we consider in our analysis and the time reversal operator $\Theta$ defined in Section \ref{sec.aps},  we have the equality $\tilde U = U$, which equivalently means that the Hamiltonian $H_\text{tot}$ is invariant under time reversal. We will now show that this implies that the work distribution of the forward process equals the work distribution of the reversed process, $p(w) = \tilde p(w)$. For each trajectory $\gamma = \{n,k,m\}$ in the forward process, there is an associated backward trajectory $\tilde \gamma = \{\tilde m,\tilde k,\tilde n \}$, but now, we will compare $\gamma$ with another trajectory belonging to the backward trajectories, that is, $\tilde \gamma ' = \{\tilde n, -\tilde k, \tilde m \}$, where if $k=ij$ is associated with the Kraus operator $M_{k=ij}$, then $-\tilde k$ corresponds to the operator $\tilde M_{ij}$ (see Eq.~(\ref{kraussHeat-tilde})). Note that if we measure the system energy at the beginning and at the end, $\ket{a_n}$ and $\ket{b_m}$ correspond to the energy eigenstates. Because the Hamiltonian is  time reversal invariant, we have that $\Theta_S \ket{a_n} = \ket{a_n}$ and $\Theta_S \ket{b_m} = \ket{b_m}$ (the same is true for the energy eigenstates of the bath, $\Theta_B \ket i = \ket i$ and $\Theta_B \ket j = \ket j$); thus, according to Eq.~(\ref{kraussHeat-tilde}), we obtain $\tilde M_{ij}=M_{ij}$. These trajectories are as follows:
\begin{equation}
\ket{n} \xrightarrow{k} \ket{m}, \quad p(\gamma) = p_i(n) \abs{\bra{b_m} M_{ij} \ket{a_n}}^2, \quad \beta w_\gamma = \ln \frac{p_i(n)}{p_i(m)} - \beta(\varepsilon_i - \varepsilon_j)
\end{equation}
\begin{equation}
\ket{\tilde n} \xrightarrow{-\tilde k} \ket{\tilde m}, \quad \tilde p(\tilde \gamma') = p_i(n) \abs{\bra{\tilde b_m} \tilde M_{ij} \ket{\tilde a_n}}^2, \quad \beta w_{\tilde \gamma'} = \ln \frac{p_i(n)}{p_i(m)} - \beta(\varepsilon_i - \varepsilon_j).
\end{equation}
Those two probabilities are the same: $p(\gamma) = \tilde p(\tilde \gamma ')$ and $w_\gamma = w_{\tilde \gamma'}$. We conclude that every trajectory in the forward process is also present in the backward process; thus, the work distribution is the same, $p(w) = \tilde p(w)$.

To derive an equivalent relation for the entropy production, $p(\Delta_i s) = \tilde p(\Delta_i s)$, we equivalently require that $\Theta\ket{a_n} = \ket{a_n}$ and $\Theta \ket{b_m} = \ket{b_m}$; however, these states are now the initial density matrix and final density matrix eigenstates. We therefore need time-reversal-invariant initial and final density matrices, i.e., $\Theta \rho \Theta^\dag = \rho$ and $\Theta \rho' \Theta^\dag = \rho'$. In the stationary state of the XX spin chain, this is fulfilled since $\rho_{XX} = e^{-\beta H_0}/Z_0$ is invariant under time reversal with $\Theta$ defined as in Section \ref{sec.aps}. However, for the XY stationary state, this is not true since $\rho_{XY} \neq \Theta \rho_{XY} \Theta^\dag$.

The previous analysis can readily be generalized to a concatenation of maps, and the same results hold.

\section{ Thermal maps do not contribute to work fluctuations}
\label{appe.thermal maps dont contribute}

To demonstrate that in the thermodynamic cycle of section \ref{sec.cycle} the work probability distribution is determined only by the driving, $p_\text{cycle}(w) = p_\text{drive}(w)$, we assume that there is just one thermalization map. The generalization to many thermalization maps, however, is straightforward.

The work distribution probability is
\begin{equation}
\begin{split}
p(w) &= \sum_\gamma p(\gamma) \, \delta(\epsilon_m - \epsilon_n - q_\gamma - w) \\
&= \sum_\gamma p_i(n)p_{i_1}p_{i_2}\,  |\bra{\epsilon_m,\varepsilon_{j_1},\varepsilon_{j_2}} U_2 U_1 \ket{\epsilon_n,\varepsilon_{i_1},\varepsilon_{i_2}}|^2 \, \delta(\epsilon_m - \epsilon_n - q_\gamma - w),
\end{split}
\end{equation}
where $U_1$ is responsible for the driving and $U_2$ is responsible for the thermalization. The probabilities are $p_i(n) = e^{-\beta \epsilon_n}/Z_S$ and $p_{i_k} = e^{-\beta \varepsilon_{i_k}}/Z_b$, and $\delta(\cdot)$ is a Kronecker-Delta. Expanding the transition probability and including two identities in the system Hilbert space $\mathcal H_S$, we obtain
\begin{equation} \nonumber
p(w) = \sum_{\gamma,\alpha,\beta} p_i(n)p_{i_1}p_{i_2} \, |\bra{\epsilon_m,\varepsilon_{j_2}} U_2 \ket{\epsilon_\alpha,\varepsilon_{i_2}} \bra{\epsilon_\alpha,\varepsilon_{j_1}} U_1 \ket{\epsilon_n,\varepsilon_{i_1}}\bra{\epsilon_n,\varepsilon_{i_1}}U_1^\dag \ket{\epsilon_\beta, \varepsilon_{j_1}} \bra{\epsilon_\beta,\varepsilon_{i_2}} U_2^\dag \ket{\epsilon_m,\varepsilon_{j_2}} \, \delta(\epsilon_m - \epsilon_n - q_\gamma - w).
\end{equation}
Since $U_2$ is thermal, $\epsilon_m + \varepsilon_{j_2} = \epsilon_\alpha + \varepsilon_{i_2} = \epsilon_\beta + \varepsilon_{i_2}$; thus, $\alpha=\beta$ because the system Hamiltonian is non-degenerate. Additionally, we can replace $\epsilon_m = \epsilon_\alpha + \varepsilon_{i_2} - \varepsilon_{j_2}$ in the delta, obtaining
\begin{equation}
p(w) = \sum_{\gamma,\alpha} p_i(n)p_{i_1}p_{i_2} \, |\bra{\epsilon_{\alpha},\varepsilon_{j_1}} U_1 \ket{\epsilon_n,\varepsilon_{i_1}}|^2 \, |\bra{\epsilon_m,\varepsilon_{j_2}} U_2 \ket{\epsilon_\alpha, \varepsilon_{i_2}}|^2 \delta(\epsilon_\alpha-\epsilon_n +\varepsilon_{j_1} - \varepsilon_{i_1}-w).
\end{equation}
Summing over $m$ and $j_2$, the second transition probability becomes a trace, which is equal to one. Finally, we arrive at
\begin{equation} \label{ecc:apenddix_work distribution cycle}
p(w) = \sum_{\alpha,j_1,n,i_1} p_i(n)p_{i_1}\, |\bra{\epsilon_\alpha,\varepsilon_{j_1}} U_1 \ket{\epsilon_n,\varepsilon_{i_1}}|^2 \delta(\epsilon_\alpha - \epsilon_n + \varepsilon_{j_1} - \varepsilon_{j_2}-w).
\end{equation}
Note that this quantity is exactly the work distribution probability of the driving alone; thus, $p_\text{cycle}(w) = p_\text{drive}(w)$. We remark that this also implies that $p_\text{cycle}(w) = \tilde p_\text{cycle}(w)$ even though the protocol is not symmetric (because in the backward process, the relaxation map acts first).

\end{appendices}

\end{document}